\documentclass[12pt]{iopart}
\usepackage{graphicx}
\usepackage{subcaption}
\usepackage{xcolor}
\usepackage{color,soul}
\expandafter\let\csname equation*\endcsname\relax

\expandafter\let\csname endequation*\endcsname\relax

\usepackage{amsmath}

\usepackage{iopams}
\begin{document}

\title[For submission in Arxiv]{Revisiting the Physics of Strong Coupling on Dusty Plasma System}

\author{J. Goswami  and S. S. Kausik }

\address{Centre of Plasma Physics, Institute for Plasma Research, Nazirakhat, Sonapur-782402, Kamrup (M), Assam, India}
\ead{kausikss@rediffmail.com}
\vspace{10pt}
\begin{indented}
\item[]November 2023
\end{indented}

\begin{abstract}
Dusty plasma can exist in a wide range, from the laboratory to Saturn's rings. The coupling parameter plays a very important role in diagnosing dusty plasma. There are two ways to deal analytically with the effect of coupling in plasma. First, where the coupling parameter is represented with the help of viscoelastic relaxation time $(\tau_m)$, and the second where the electrostatic temperature $(d)$ represents the coupling parameters, in this work, we have used both techniques in a strongly coupled four-component plasma in which we have considered the dynamics of negatively charged dust grains and $q-$ non-extensive distribution for electrons, ions, and positrons. The main focus of this particular work is to find out the operational regions and validity of these two techniques. The result of this paper will give future researchers an idea of how coupling would be encountered in a dusty plasma system.
\end{abstract}

%
%
%
%
%

\section{Introduction:}\label{sec1}
Dust grains in any plasma increase the complexity of the plasma, which is why dusty plasma is most popularly known as `complex plasma'\cite{shukla2001survey}. Dusty or complex plasma can present in various astrophysical bodies \cite{Goertz,Bliokh,Shukla2,Mendis199553,Mendis2,shukla3}, namely, in the planetary rings, in interstellar molecular clouds, in cometary comae and tails, etc. Dust particles are also very much present in different types of nebulae. On the other hand, dusty or complex plasma environments can also be created in the laboratory. Polar mesosphere summer echoes (PMSE) can also be explained by the presence of charged dust particles there\cite{shukla2001survey,Cho,ZHOU1997739}. There are various physical processes like photoelectron emission, secondary electron emission, thermionic emission, etc., by which dust grains can be negatively or positively charged\cite{Mott-Smith,sodha1971advances,Rosenberg376584,Rosenberg763125,Whipple,Paul_2022}. 
\par
Generally, astrophysical plasmas in Earth's bow shock\cite{asbridge1968outward}, the vicinity of the moon \cite{futaana2003moon}, Jupiter's magnetosphere\cite{horanyi1993mechanism}, Saturn's magnetosphere\cite{krimigis1983general,horanyi2004dusty}, galaxy clusters \cite{hansen2005cluster} , etc exhibit the long-range interactions. Therefore, these systems do not obey the extensive statistics but follow the statistics theoretically proposed by Tsallis\cite{tsallis1988possible}. The distribution following the above-mentioned statistics is also observed by the Voyager I space mission \cite{goswami2018study}.
\par
Electron–positron–ion-dust (EPID) plasma has been widely found in various astrophysical bodies like Jupiter’s magnetosphere\cite{horanyi1993mechanism}, supernova environments\cite{pilipp19813}, Saturn’s magnetosphere\cite{krimigis1983general,horanyi2004dusty}, the solar atmosphere\cite{horanyi1996charged,murphy2005physics}, galactic centre \cite{zurek1985annihilation}, and cometary tails\cite{horanyi1996charged,ellis1991numerical,mendis1994cosmic}. There are three main mechanisms viz. pair production\cite{murphy2005physics}, thermal heating\cite{nosenko2008heat,avinash2011anomalous}, and radiative heating \cite{rosenberg1999positively} by which charged dust species or positron or both occur naturally in electron–ion plasma. The light positron and the heavy dust grains very prominently control the dynamics of the EPID plasma medium (EPIDPM). Electrostatic potential structure\cite{banerjee2016arbitrary}, the existence of the positive super solitons \cite{paul2016dust}, and the stability of the system \cite{rahman2021dust} have been discussed by various authors in this type of four-component plasma.
\par
In this work, we are mainly focusing on the effect of coupling in the dusty plasma. Dusty plasma can be divided into two varieties in light of coupling, weakly coupled and strongly coupled dusty plasma. Theoretically, this difference can be depicted by the coulomb coupling parameter,
\begin{equation}
    \Gamma=\frac{Q^2}{dT_d}exp(-\kappa)
\end{equation}
Where $Q=Z_{d0}e$; is the charge on the surface of the dust grains, $T_d$ is the temperature of dust, and $\kappa=\frac{d}{\lambda_d}$; $d$ is an average intergrain distance and $\lambda_d$ is the dusty plasma Debye radius. Thus, $\Gamma \le 1$ represents a weakly coupled plasma and $\Gamma > 1$ represents a strongly coupled plasma. Again, if $\Gamma > \Gamma_c$ (a critical value of the coupling parameter) the dust grains form a crystalline structure. In the general hydrodynamic model of dusty plasma, there are two ways to encounter the problem of strong coupling,
\par
a) when the coupling is represented by normalized viscoelastic relaxation time \cite{MAMUN2004412,el2012dust}
\par
b) when the coupling is represented by the normalized electrostatic temperature \cite{Cousens}. 
\par
Various authors have used these two methodologies to handle the coupling individually, but no one has ever done a comparative discussion of these two. So, in this article, we basically want to address the following questions for the first time:
\par
a) Are there any connections between these two types of operations for strong coupling?
\par
b) If the answer to the previous question is no then why? But if the answer is yes then what are the connections?
\par
c) and lastly can we use any of these two procedures for all the plasma systems? Are there any advantages or disadvantages of these models? 
\par
Here, we mainly focus on the four component dusty plasma present in Saturn's magnetosphere. But for simplicity, we have not taken the effect of the magnetic field ( of strength $\sim 0.025 - 0.21 G$ \cite{angeo-24-1145-2006}) otherwise it will become a 3D problem.\cite{Goswamiradiation}. 
\par
There are two types of dissipation in the plasma system, collision, and viscosity\cite{Goswami_2023}. These dissipative forces come to act in the momentum equations. Three situations can arise with the presence of dissipation. If the dissipation is moderate then we have oscillatory shock waves. Monotonic shock waves can be found with a higher value of dissipation, and if there is no dissipation present in the system then solitary waves will be propagated in the system. But, in an astrophysical scenario, the viscous co-efficient will have a dominant effect over the collisional term \cite{Sulaiman2016}.
\par
\par
In summary, we can say that we are focusing on viscous collisionless astrophysical EPID plasma where we discuss for the first time two procedures to deal with the strong coupling in the same plasma model. From these calculations, we are able to conclude for the maiden time which operation is suitable for analyzing the linear and non-linear characteristics of dust acoustic (DA) waves. As we are working on dust acoustic (DA) mode, only the motion of dust is considered to construct the model and the rest of the quantities that is ions, electrons, and positrons have been $q-$nonextensive distributions. 
\par
The paper is arranged in the following way- in section \ref{sec2} we have formulated the problem considering strongly coupled dust and this coupling is represented by normalized viscoelastic relaxation time. Then we deal with the linear and nonlinear characteristics of the same model in section \ref{sec3}. In section \ref{sec4}, we have analyzed the model for different coupling conditions. Then section \ref{sec5} we have discussed some of the important results. Lastly, in section \ref{sec6} we have concluded with important findings from the work and the prospect.

\section{Basic Formulations:}\label{sec2}
We have considered a four-component electron-ion-positron-dust plasma. To investigate the dust acoustic mode we have to consider the normalized dynamical equation of dust only. The continuity equation for the dust grains is given by,
\begin{equation}\label{eq1}
    \frac{\partial n_d}{\partial t}+\frac{\partial}{\partial x}\left(n_du_d\right)=0
\end{equation}
The momentum equation for strongly coupled dust is given by,
\begin{equation}\label{eq2}
    \left(1+\tau_m D_t\right)\left[n_d\left(D_t u_d - \frac{\partial \varphi}{\partial x}\right)\right]=\eta_d\frac{\partial^2 u_d}{\partial x^2}
\end{equation}
Poisson's equation bounds these two equations:
\begin{equation}\label{eq3}
    \frac{\partial^2 \varphi}{\partial x^2}=n_d +\alpha_e n_e-\alpha_i n_i-\alpha_p n_p
\end{equation}
These equations (\ref{eq1}-\ref{eq3}) are normalized by standard normalization scheme and these are, $\hat{x} \rightarrow \frac{x}{\lambda_D}$, $\hat{t} \rightarrow \omega_p t$, $\hat{\varphi} \rightarrow \frac{e\varphi}{T_d}$, $\hat{u_d} \rightarrow \frac{u_d}{C_d}$, where, $C_d=\left(\frac{Z_d T_d}{m_d}\right)^{1/2}$, ${\omega_p}^{-1}=\left(\frac{m_d}{4\pi Z_d e^2 n_{d0}}\right)^{1/2}$ and $\lambda_D=\left(\frac{T_d}{4\pi Z_d n_{d0}e^2}\right)^{1/2}$. Again, $\eta_d$ is the normalized longitudinal viscous coefficient, and $\tau_m$ is the normalized viscoelastic relaxation time normalized by dust time period $(\tau_d)$ \cite{MAMUN2004412,el2012dust}.
\par
We consider the $q-$nonextensive distribution for electrons, positrons, and ions. The normalized number densities of these are given below,
\begin{equation}\label{eq4}
    n_e=\left[1+\left(q_e-1\right)\sigma_e\varphi\right]^{\frac{1+q_e}{2\left(q_e-1\right)}}
\end{equation}
\begin{equation}\label{eq5}
    n_p=\left[1-\left(q_p-1\right)\sigma_p\varphi\right]^{\frac{1+q_p}{2\left(q_p-1\right)}}
\end{equation}
\begin{equation}\label{eq6}
    n_i=\left[1-\left(q_i-1\right)\sigma_i\varphi\right]^{\frac{1+q_i}{2\left(q_i-1\right)}}
\end{equation}
where, $\sigma_j=\frac{T_j}{T_d}$; $(j=e.p,i)$. 
\par
All the species have different $q-$values in the model. Basically, the degree of non-extensivity is measured by this non-extensive parameter $q$. If $q=1$ then the distribution is transformed into Maxwellian distribution. Also, $q<1$ and $q>1$ represent super-extensive and sub-extensive distributions, respectively. We considered super extensive distribution for all the species and for convenience we have consider the same values of $q-$ parameters in the section \ref{sec5}. Dust charging frequency is $10^4 - 10^5$ times higher than the frequency of dust acoustic waves(DAWs) in this type of plasma. Therefore, in our derivation, we have ignored the effect of dust charge fluctuations.
\par
One more thing worth mentioning here is the Debye radius of dusty plasma($\lambda_D$). As we consider the model as an electron-positron-ion-dust (EPID) model then the $\lambda_D$ can be expressed as below \cite{shukla2001survey},
\begin{equation}
    \frac{1}{{\lambda_D}^2}=\frac{1}{{\lambda_{De}}^2}+\frac{1}{{\lambda_{Dp}}^2}+\frac{1}{{\lambda_{Di}}^2}
\end{equation}
where $\lambda_{De}=\left(\frac{T_e}{4\pi n_{e0} e^2}\right)^{1/2}$, $\lambda_{Dp}=\left(\frac{T_p}{4\pi n_{p0} e^2}\right)^{1/2}$ and $\lambda_{Di}=\left(\frac{T_i}{4\pi n_{i0} e^2}\right)^{1/2}$ are the Debye radius of electrons, protons, and ions, respectively. Here, $T_e$, $T_p$, and $T_i$ are the temperatures of the species. Also, $n_{e0}$, $n_{p0}$, and $n_{i0}$ are the unperturbed electrons, positrons, and ions number density respectively. Where electrons and positrons are created by pair production there all the parameters except the charge are the same for these two species. So if the temperatures and densities of electrons and ions are almost equal then the Debye radii are also equivalent to each other. But we are working in a region where $Tp=Te>>T_i$ and ion density is greater than the density of the electrons and positrons. So in our case $\lambda_D \sim \lambda_{Di}>>\lambda_{De}$.
\section{Linear and Nonlinear Analysis:}\label{sec3}
For the linear and nonlinear analysis of the dust acoustic mode in four component electron-positron-ion-dust plasma we consider the reductive perturbation of the field quantities $n_d$, $u_d$, and $\varphi$.
\begin{equation}\label{eq7}
    n_d=1+\sum_{m=1}^\infty \epsilon^m {n_d}^{(m)} 
\end{equation}
\begin{equation}\label{eq8}
    u_d=u_0+\sum_{m=1}^\infty \epsilon^m {u_d}^{(m)} 
\end{equation}
and,
\begin{equation}\label{eq9}
    \varphi=\sum_{m=1}^\infty \epsilon^m {\varphi}^{(m)} 
\end{equation}
\subsection{Linear Dispersion Characteristics:}\label{sec3.1}
In order to derive the linear dispersion characteristics we assume all the field variables are varying as $exp i(kx-\omega t)$, now using the equations (\ref{eq7}-\ref{eq9}) we get,
\begin{equation}\label{eq10}
    a\omega^3+b\omega^2+c\omega+d=0
\end{equation}
The constants $a,b,c$ and $d$ are given by,
\begin{equation}\label{eq11}
\left. \begin{array}{ll}
\displaystyle a=\left(k^2 \tau_m+\tau_m D\right)
  \\[8pt]
  \displaystyle b=\left(k^2+D\right)\left[\tau_m k u_0 -1\right] \\[8pt]
  \displaystyle c=\left(k^2+D\right)\left[2ku_0 - \tau_m u_0 k^2-2\tau_m k^2 {u_0}^2-k^2\eta_d\right]+\tau_m k^2 \\[8pt]
  \displaystyle d=\left(k^2+D\right)\left[k^3 u_0 \eta_d + \tau_m {u_0}^2 k^3-k^2{u_0}^2\right]+k^2-\tau_m u_0 k^3 \\[8pt]
 \end{array}\right\}
\end{equation}
where, $D$ in the equation \ref{eq11} is
\begin{equation}\label{eq12}
    D=\frac{{\alpha_e}^2 \sigma_e}{2}(1+q_e)+\frac{{\alpha_i}^2 \sigma_i}{2}(1+q_i)+\frac{{\alpha_p}^2 \sigma_p}{2}(1+q_p)
\end{equation}
Clearly, the dispersion relation has three roots, by solving the equation (\ref{eq10}) we get, 
\begin{equation}\label{eq13}
\left. \begin{array}{ll}
\displaystyle \omega_1=S+T-\frac{b}{3a}
  \\[8pt]
  \displaystyle \omega_2=-\frac{S+T}{2}-\frac{b}{3a}+\frac{i\sqrt{3}}{2}(S-T) \\[8pt]
  \displaystyle \omega_3=-\frac{S+T}{2}-\frac{b}{3a}-\frac{i\sqrt{3}}{2}(S-T) \\[8pt]
 \end{array}\right\}
\end{equation}
A full solution with the values of $S$ and $T$ is given in the \ref{A}.

\subsection{KdVB and KdV Equations:}\label{sec3.2}

In order to investigate the nonlinear behavior of dust acoustic waves the reductive perturbation techniques of the field quantities given in the equations (\ref{eq7}-\ref{eq9}) have been used. Also, we use the stretching of space, time variables, and viscous, coupling parameters as follows,
\begin{equation}\label{eq14}
\left. \begin{array}{ll}
\displaystyle \xi=\epsilon^{1/2}\left(x-Mt\right)
  \\[8pt]
  \displaystyle \tau=\epsilon^{3/2}t \\[8pt]
  \displaystyle \tau_m=\epsilon^{1/2}\tau_0 \\[8pt]
  \displaystyle \eta_d=\epsilon^{1/2}\eta_0 \\[8pt]
 \end{array}\right\}
\end{equation}
Where, $M$ is the Mach number and $\epsilon$ is the strength of the nonlinearity $(0<\epsilon<1)$.
\par
We yield the KdV-Burgers' equation as follows,
\begin{equation}\label{eq15}
    \frac{\partial \varphi^{(1)}}{\partial \tau}+A\varphi^{(1)}\frac{\partial \varphi^{(1)}}{\partial \xi}+B\frac{\partial^3 \varphi^{(1)}}{\partial \xi^3}-C\frac{\partial^2 \varphi^{(1)}}{\partial \xi^2}=0
\end{equation}
Where, $A$, $B$, and $C$ are the nonlinear, dispersive, and dissipative coefficients, respectively. They are given below
\begin{equation}\label{eq16}
\left. \begin{array}{ll}
\displaystyle A=B\Bigg[ \sum_{j=i,p,e} (\pm)_j \alpha_j\sigma_j^2 \frac{(1-q_j)(3-q_j)}{8}-\frac{3}{\left(M-u_0\right)^4}\Bigg]
  \\[8pt]
  \displaystyle B=\frac{(M-u_0)^3}{2} \\[8pt]
  \displaystyle C=\frac{\eta_0}{2} \\[8pt]
 \end{array}\right\}
\end{equation}
The sign of $A$ will be $+$ for $j=i$ and $p$ and $-$ for $j=e$.
\par
This KdVB equation (\ref{eq15}) has no known exact solution but there are some special solutions to this equation. The dissipative coefficient is only dependent on the viscous coefficient which means if the viscosity of the medium increases, the dissipative nature of the system also increases. The dispersive coefficient is depending on the Mach number and the streaming velocity of the dust particles. The nonlinear term depends on $M$, $u_0$, $\alpha_j$, $\sigma_j$ , and $q_j$ $(j=i,p$ and $e$).
\par
Following the procedure given by Pakzad and Zavidan \cite{pakzad2009dust} we derived two solutions to the equation:
\par
\space a. For $C^2<<4BM$ we get the oscillatory shock wave (OSW) solution given by,
\begin{equation}\label{eq17}
    \varphi^{(1)}=\varphi_c+\varphi_0 exp\left(\frac{C}{2B}\psi\right)cos\sqrt{\frac{M}{B}}\psi
\end{equation}
\par
\space b. For $C^2>4BM$ we get the monotonic shock wave (MSW) solution given by,
\begin{equation}\label{eq18}
    \varphi^{(1)}=\frac{12A}{B}\left[1-tanh^2\psi\right]-\frac{36C}{15A}tanh \psi
\end{equation}
Clearly, If there is no dissipation that is no viscous effect $(\eta_0=0)$ in the present case, then the KdVB equation (\ref{eq15}) is transformed into the KdV equation,
\begin{equation}\label{eq19}
    \frac{\partial \varphi^{(1)}}{\partial \tau}+A\varphi^{(1)}\frac{\partial \varphi^{(1)}}{\partial \xi}+B\frac{\partial^3 \varphi^{(1)}}{\partial \xi^3}=0
\end{equation}
The solution of this KdV equation would give us the solitary structures given by,
\begin{equation}\label{eq20}
    \varphi^{(1)}=\frac{3M}{A}sech^2 \left(\frac{\psi}{2\sqrt{\frac{M}{B}}}\right) 
\end{equation}
\section{Coupling in the liquid-crystal region:}\label{sec4}
\subsection{Reason of considering another process:}\label{sec4.1}
The effect of coupling is very important for the description of the crystallization of a dusty plasma system. In fact, it is the main reason for forming this type of ordered structure in a dusty plasma medium. But, from the equations (\ref{eq15}) and (\ref{eq16}) it can be clearly observed that there is no visible effect of the viscoelastic relaxation time in the nonlinear KdVB equation. So there may be two possibilities,
\par
a. Coupling does not have any effect on the nonlinear structures.
\par
b. The effect of coupling is absorbed by the other variables and coefficients.
\par
If we consider the first reason is true then the model given above will be valid in a very limited region. But various authors \cite{Kaw1} have shown that models like the above mentioned are valid at least up to $\Gamma \sim \Gamma_c$. So coupling must have some effect on the nonlinear structures too. From the work of previous scientists \cite{Kaw1} we can write the expression of the viscoelastic relaxation time as
\begin{equation*}
   \tau_m = \eta_d \frac{T_e}{T_d}\left[-0.396 \Gamma + 0.739 \Gamma^{\frac{1}{4}} + 0.058 \Gamma^{-\frac{1}{4}}-0.27\right]^{-1} 
\end{equation*}
From this equation, it is very clear that the viscous coefficient has a direct relation with $\tau_m$ so the second reason is the correct one. Also as we have considered the stretching of the space and time-like variables along with the viscoelastic relaxation time that is why the effect of the coupling coefficients may have been absorbed by the other variables. So to find out the effect of the coupling individually of the system with higher $\Gamma$ i.e., near the liquid crystal region, we have to treat the coupling in a different way\cite{Cousens}. 

\subsection{Liquid-crystal Model:}
For the reason mentioned in the section \ref{sec4.1} we only have to change the momentum equation of the governing equation, and the new equation is given by,
\begin{equation}\label{eq25}
n_d\Bigg(\frac{\partial u_d}{\partial t}+u_d\frac{\partial u_d}{\partial x} \Bigg)=n_d\frac{\partial \varphi}{\partial x}-\frac{\partial}{\partial x}\left(n_d d\right)+\eta_d \frac{\partial^2 u_d}{\partial x^2}    
\end{equation}
Here, the second term in the right-hand side of the equation (\ref{eq25}) is the representation of strong coupling. Before starting any operation with the help of this equation we have to understand the factor `$d$'. This $d$ can be expressed as,
\begin{equation}\label{eq26}
    d=d_0+\epsilon d_1+\epsilon^2 d_2+...
\end{equation}
where,
\begin{equation}\label{eq27}
\left. \begin{array}{ll}
\displaystyle d_1 = d_{11}n^{(1)}+d_{12}\varphi^{(1)}
  \\[8pt]
  \displaystyle d_2=d_{21}n^{(2)}+d_{22}\varphi^{(2)}+d_{23}{n^{(1)}}^2+d_{24}n^{(1)}\varphi^{(1)}+d_{25}{\varphi^{(1)}}^2\\[8pt]
 \end{array}\right\}
\end{equation}
Here,
\begin{equation}\label{eq28}
\left. \begin{array}{ll}
\displaystyle d_{11}=d_{21}=\frac{d_0}{3}\frac{1+\kappa_0+{\kappa_0}^2}{1+\kappa_0}
  \\[8pt]
  \displaystyle d_{12}=d_{22}=-d_0 c_2 \frac{{\kappa_0}^2}{1+\kappa_0}\\[8pt]
  \displaystyle d_{23}=\frac{d_0}{18}\frac{{\kappa_0}^3-3{\kappa_0}^2-2\kappa_0-2}{1+\kappa_0}\\[8pt]
  \displaystyle d_{24}=-\frac{d_0}{3}c_2\frac{{\kappa_0}^2 \left(\kappa_0-1\right)}{1+\kappa_0}\\[8pt]
  \displaystyle d_{25}=-\frac{d_0}{2}\left(3c_3-{c_2}^2 \kappa_0 \right)\frac{{\kappa_0}^2}{1+\kappa_0}\\[8pt]
 \end{array}\right\}
\end{equation}
The values of $c_2$ and $c_3$ are given in the \ref{B}. And, the $d$ is the normalized electrostatic temperature and $\kappa_0$ is a dimensionless quantity which is the ratio of inter-particle distance and the Debye length. This $\kappa_0$ is also the measurement of the coupling strength in the plasma. 
\par
As we mentioned above a model of this type was first developed to define the crystalline plasma structures but nowadays such models have a wide range of uses in dusty plasma. We can also say that this is the most general model for a dusty plasma and will be shown in our discussion in the section \ref{sec5}. Immediately one question may arise why do we use the previous model? In section \ref{sec5} we will answer this question too.
\par
At first, we calculated the linear dispersion relation for this type of mathematical representation of the coupling effect. We get the dispersion relation as,
\begin{equation}\label{eq29}
    a_1 \omega^2+b_1 \omega +c_1 = 0
\end{equation}
Here,
\begin{equation}\label{eq30}
    \left. \begin{array}{ll}
    \displaystyle a_1 = \left(k^2+D\right) \\[8pt]
    \displaystyle b_1 = - \left(2ku_0+k^2 \eta_0\right)a_1\\[8pt]
    \displaystyle c_1 = a_1 \left(k^2 {u_0}^2 + k^3 \eta_0 u_0 -k^2 \beta\right)-\alpha k^2\\[8pt]
    \end{array}\right\}
\end{equation}
where, $D$ is previously given in equation (\ref{eq12}) and $\alpha$ \& $\beta$ is given below,
\begin{equation}\label{eq32}
    \left. \begin{array}{ll}
    \displaystyle \alpha = 1- d_{12} \\[8pt]
    \displaystyle \beta = d_0+d_{11}\\[8pt]
    \end{array}\right\}
\end{equation}
Solving the equation \ref{eq29} we get,
\begin{equation}\label{eq31}
  \left. \begin{array}{ll}
    \displaystyle {\omega_1}' = \frac{-b_1+\sqrt{{b_1}^2-4a_1 c_1}}{2a_1} \\[8pt]
    \displaystyle {\omega_2}' = \frac{-b_1-\sqrt{{b_1}^2-4a_1 c_1}}{2a_1}\\[8pt]
    \end{array}\right\}  
\end{equation}
The most important thing from this calculation is the effect of coupling in the nonlinear regime. So with the help of the previously used reductive perturbation technique with the newly used equation (\ref{eq26}) we get another KdVB equation,
\begin{equation}\label{eq33}
    \frac{\partial \varphi^{(1)}}{\partial \tau}+A_1 \varphi^{(1)} \frac{\partial \varphi^{(1)}}{\partial \xi}+B_1 \frac{\partial^3 \varphi^{(1}}{\partial \xi^3}-C_1 \frac{\partial^2 \varphi^{(1)}}{\partial \xi^2}=0
\end{equation}
Here, 
\begin{equation}\label{eq34}
    \left. \begin{array}{ll}
    \displaystyle A_1 = \frac{-\Lambda \left(\alpha-2d_{12}\alpha-2d_{24}\alpha-2d_{25}\Lambda\right)-2\left(d_{11}+d_{23}\right)\alpha^2+2 \alpha^2 \left(M-u_0\right)^2 \Lambda+2c_2 \Lambda^3}{2\alpha \left(M-u_0\right) \Lambda} \\[8pt]
    \displaystyle B_1 = \frac{\Lambda^2}{2\alpha \left(M-u_0\right)} \\[8pt]
    \displaystyle C_1 = \frac{\eta_0}{2 \Lambda}\\[8pt]
    \end{array}\right\}
\end{equation}
where, $\Lambda=\left[\beta-\left(M-u_0\right)^2\right]$ and other quantities are given in the equations (\ref{eq28}) and (\ref{eq32}). The solution of equation (\ref{eq33}) will be the same as given in the equations (\ref{eq17}) and (\ref{eq18}).
\section{Result:}\label{sec5}
In this paper, we mainly take physically acceptable values of plasma parameters for astrophysical plasma systems. The values of $M$, $u_0$, $\eta_0$, $q$ and $\kappa_0$ have given in the table \ref{table1}.
\begin{table}[!h]
\begin{center}
\begin{tabular}{||c c c c||} 
 \hline
 Plasma Parameters & Range(Normalized) & Plasma & References \\ [0.5ex] 
 \hline\hline
 $M$ & $1.2$-$3.5$ & Saturn's Orbit & \cite{ECHER2019210} \\ 
 \hline
 $u_0$ & $0.01$-$0.6$ & Laboratory & \cite{Merlino,Ishihara20} \\
 \hline
 $\eta_0$ & $0.15$-$0.55$ \& $3$-$5$ & Theoretical \& Astrophysical & \cite{el2012dust,Goswami_2023,pakzad2009dust} \\
 \hline
 $q$ & $0.459$-$0.846$ & Saturnian MSp $e^-$, $H^+$ \& $O^+$ & \cite{dialynas2009energetic,schippers2008multi} \\
 \hline
 $\kappa_0$ & $0.97$, $3.22$, $5.80$ &  White dwarf, Jovian interior & \cite{Schatzman,salpeter1973convection,Ichimaru,Vaulina2000} \\ [1ex] 
 \hline
\end{tabular}
\caption{\label{table1}Important Plasma Parameters for various Dusty Plasma Systems}
\end{center}
\end{table}

In this section, we will mainly discuss the effects of two types of treatments of the coupling effect. With the help of the calculations in the previous sections and from the figures here, we will try to answer all the questions we have raised in the introduction section [section \ref{sec1}]. 
\par
Before going to that part we will first discuss two things
\par
a) In which equation did the main difference between the two techniques incorporate?
\par
b) Why have the coupling of electrons and ions not been taken?
\par
To answer the first question we can say that the main difference between these two treatments is in the momentum equations given in equations (\ref{eq2}) and (\ref{eq25}). In both cases, we have considered the electrons and ions to be weakly coupled as their temperatures are very high compared to the dust grains. 
\par
Now we will try to discuss the answer to the first question of the introduction. The model with the equation (\ref{eq2}) was developed to analyze the plasma system where plasma is treated as a fluid. The second treatment of the coupling was initially developed for crystalline plasma structure but later it was also used in the fluid section [ equation \ref{eq25}]. In the next two subsections \ref{sec5.1} and \ref{sec5.2} we will discuss the linear and nonlinear waves respectively for the two types of treatments. 
\par
\subsection{Discussions on Linear waves:}\label{sec5.1}
\begin{figure}
\begin{subfigure}{.32\textwidth}
  \centering
  \includegraphics[width=1\linewidth]{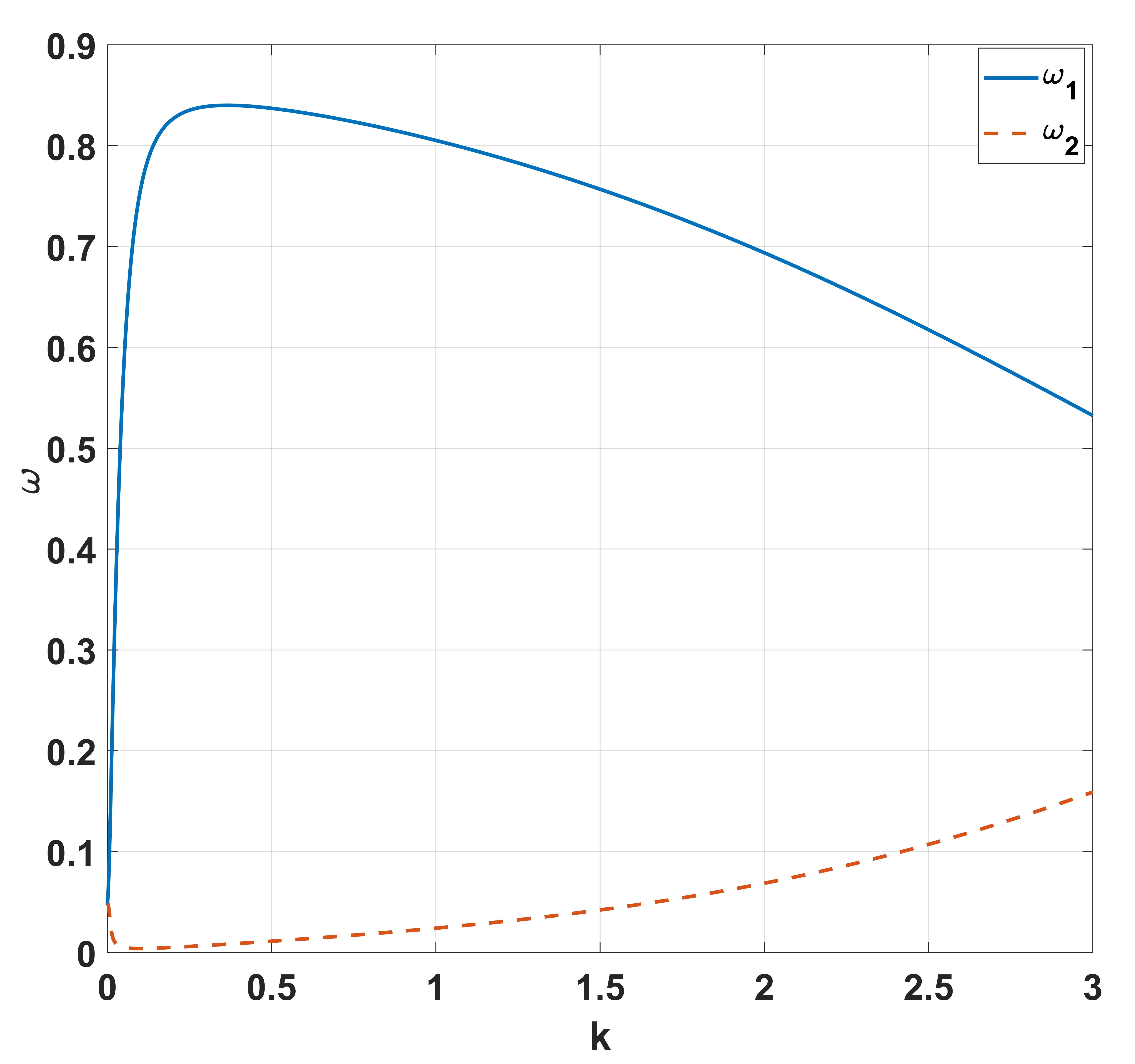}  
  \caption{}
  \label{dispersionrelationoffirstkindforroots230523}
\end{subfigure}
\begin{subfigure}{.32\textwidth}
  \centering
  \includegraphics[width=.9\linewidth]{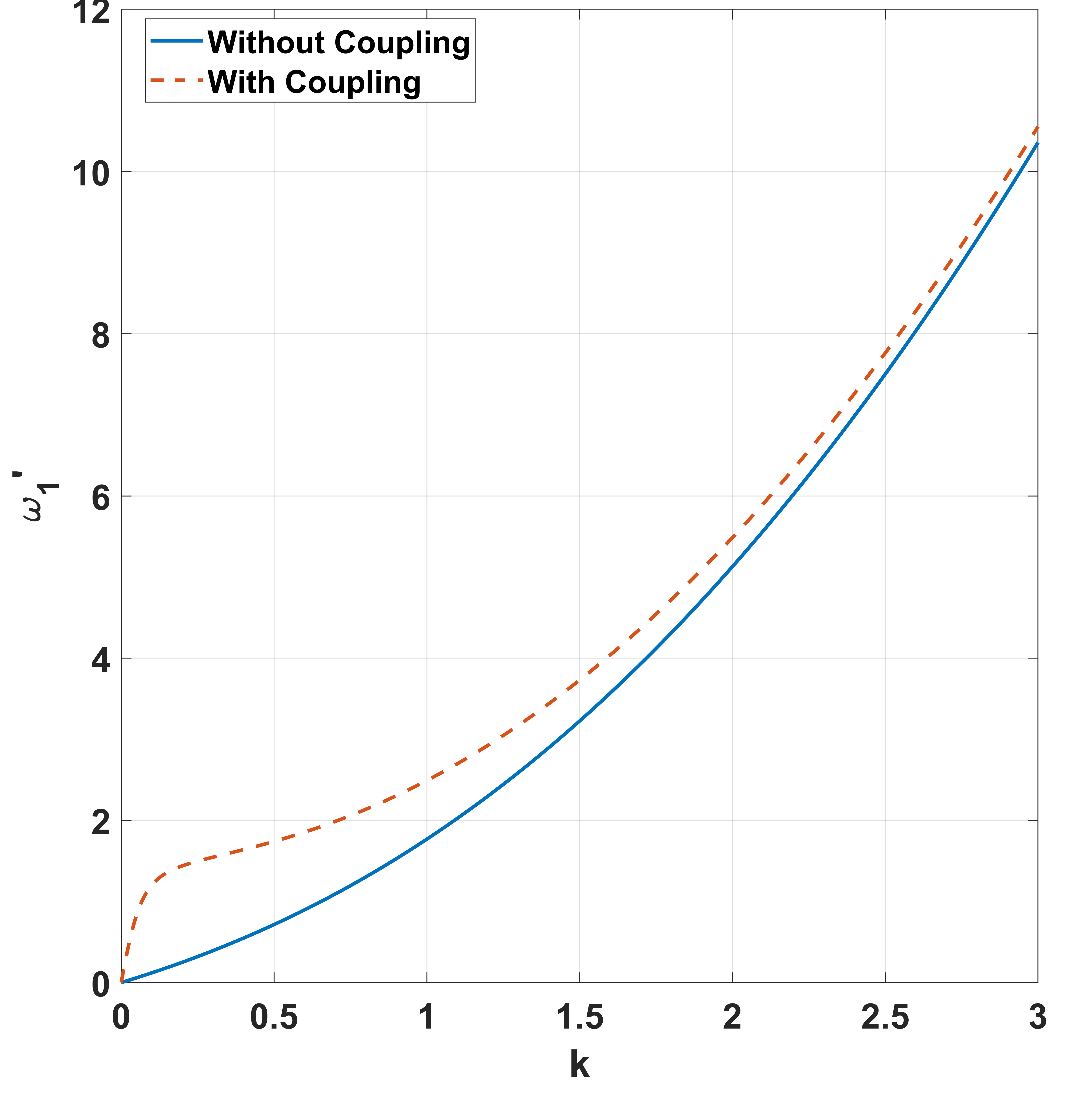}  
  \caption{}
  \label{Dispersionofsecondkindwithandwithoutcoupling}
\end{subfigure}
\begin{subfigure}{.32\textwidth}
  \centering
  \includegraphics[width=.9\linewidth]{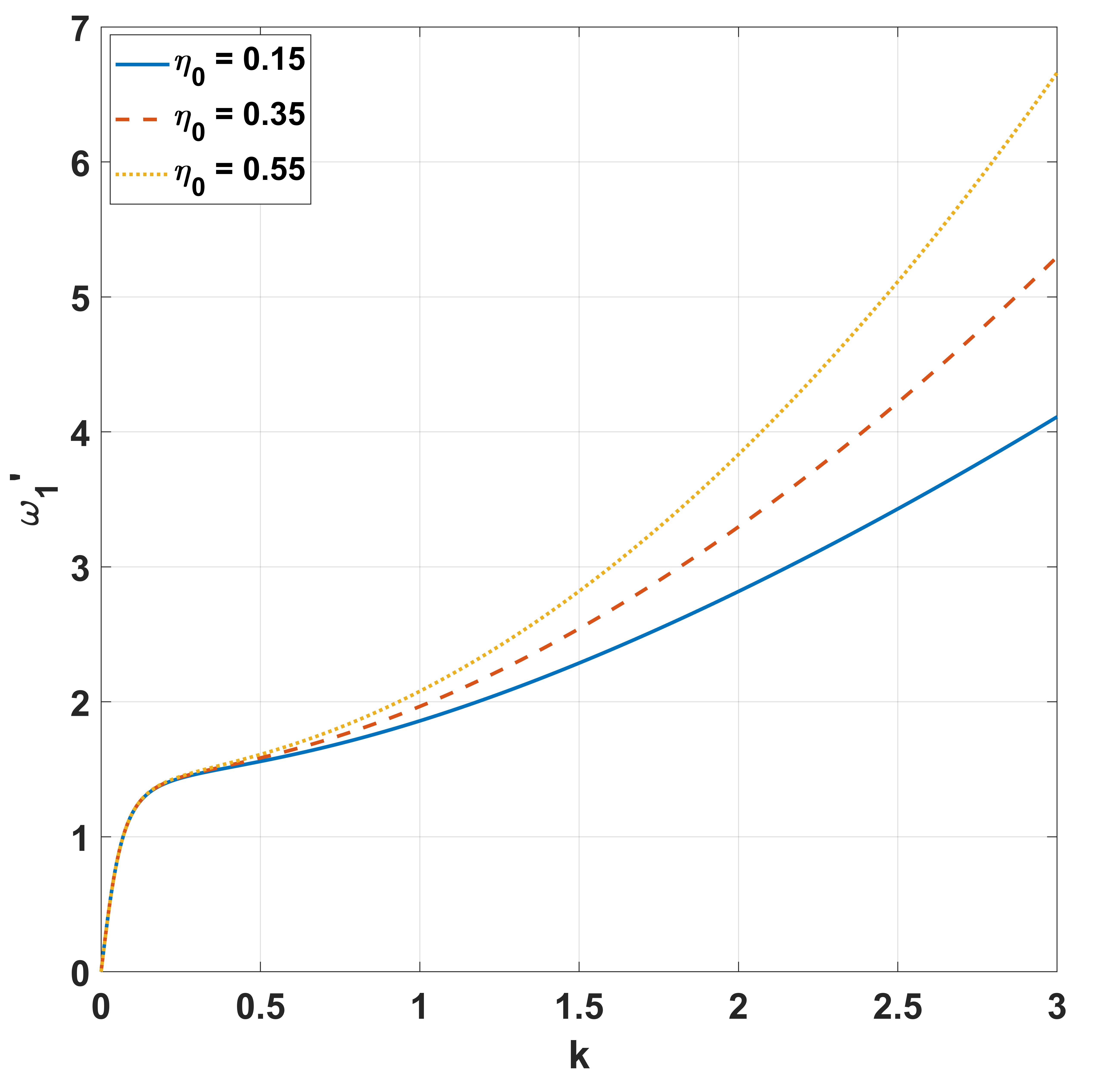}  
  \caption{}
  \label{Dispersionofsecondkindfordifferenteta0}
\end{subfigure}
\caption{(a) Two roots of the dispersion relation (\ref{eq10}) (b) Dispersion relation plot from equation (\ref{eq29}) with and without coupling (c) Dispersion relation plot from equation (\ref{eq29}) with different values of the dust longitudinal viscosity.}
\label{fig1}
\end{figure}

\begin{figure}
\begin{subfigure}{.32\textwidth}
  \centering
  \includegraphics[width=.9\linewidth]{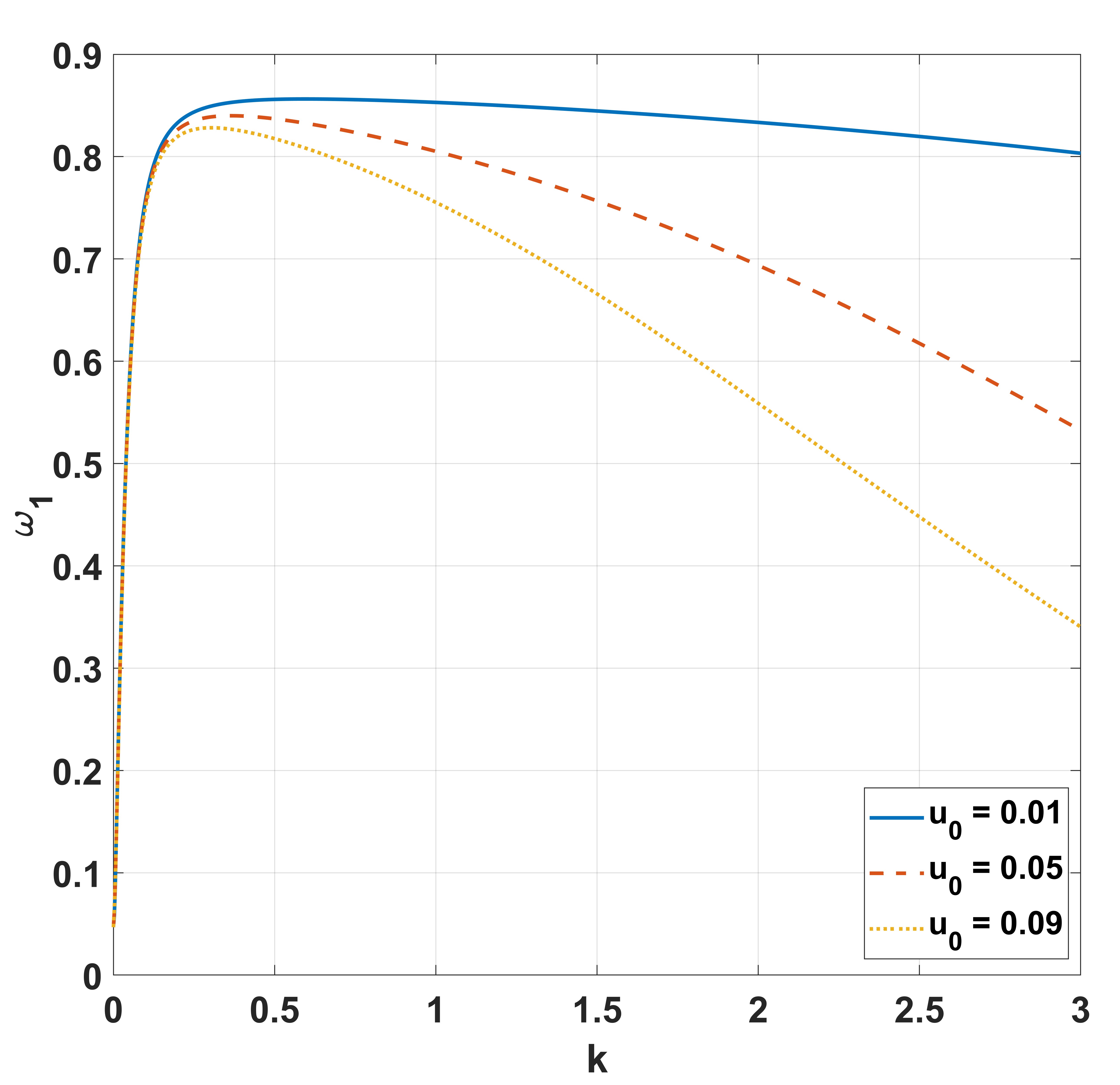}  
  \caption{}
  \label{dispersionrelationoffirstkindfordifferentu0230523}
\end{subfigure}
\begin{subfigure}{.32\textwidth}
  \centering
  \includegraphics[width=.9\linewidth]{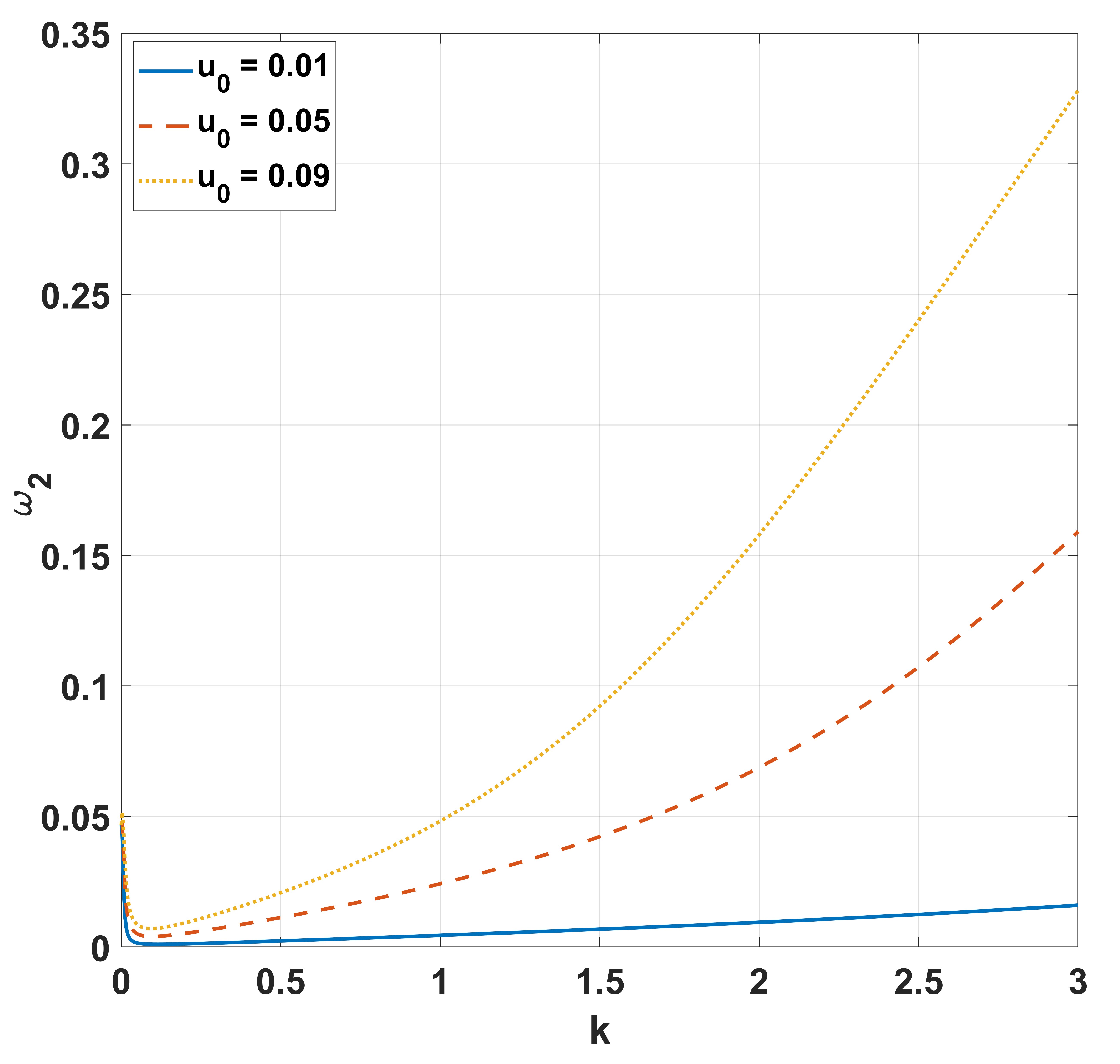}  
  \caption{}
  \label{dispersionrelationoffirstkindfordifferentu02230523}
\end{subfigure}
\begin{subfigure}{.32\textwidth}
  \centering
  \includegraphics[width=.9\linewidth]{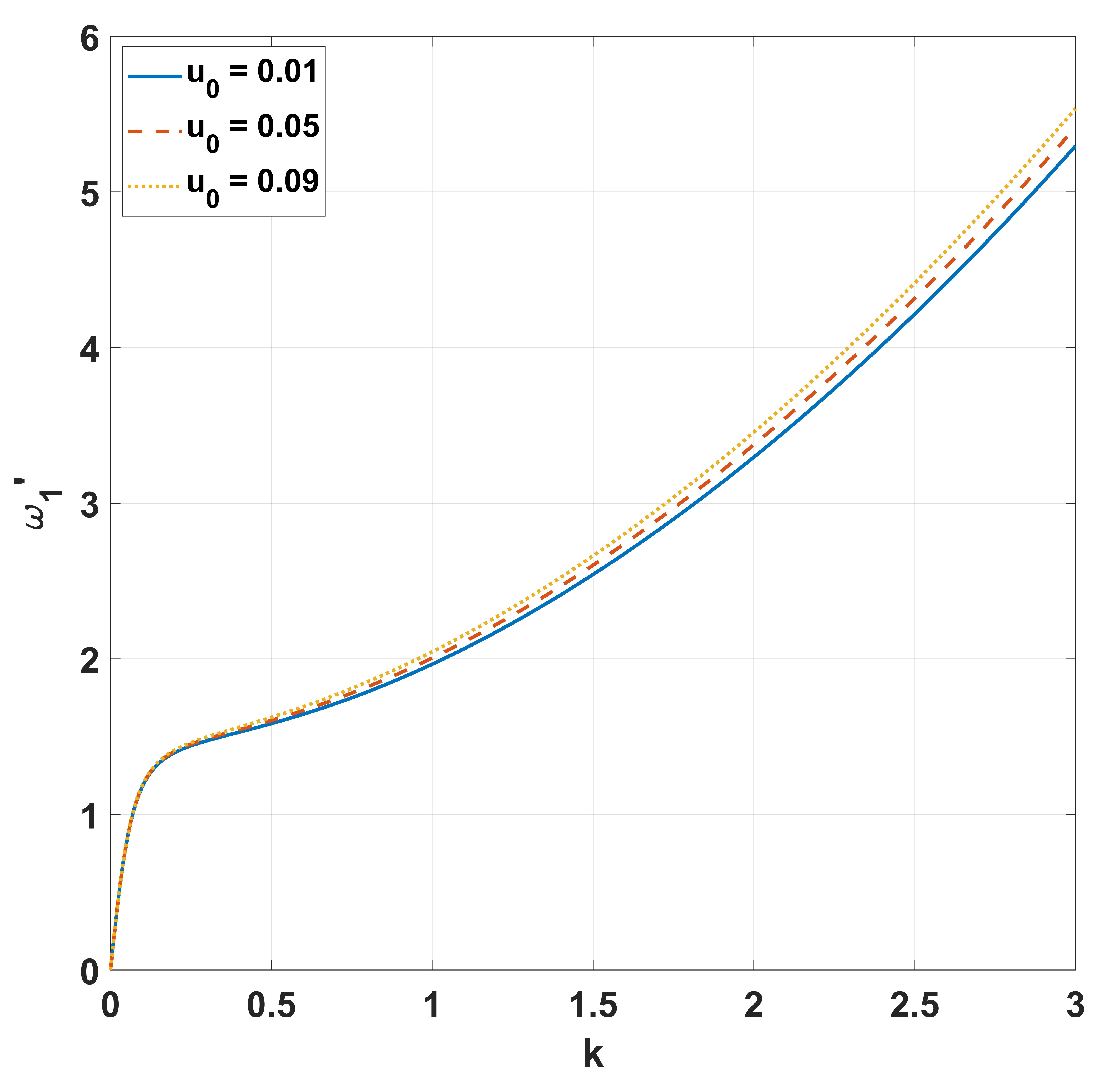}  
  \caption{}
  \label{Dispersionofsecondkindfordifferentuu0}
\end{subfigure}
\caption{Dispersion relation for different streaming velocities of dust}
\label{fig2}
\end{figure}

\begin{figure}
\begin{subfigure}{.5\textwidth}
  \centering
  \includegraphics[width=.85\linewidth]{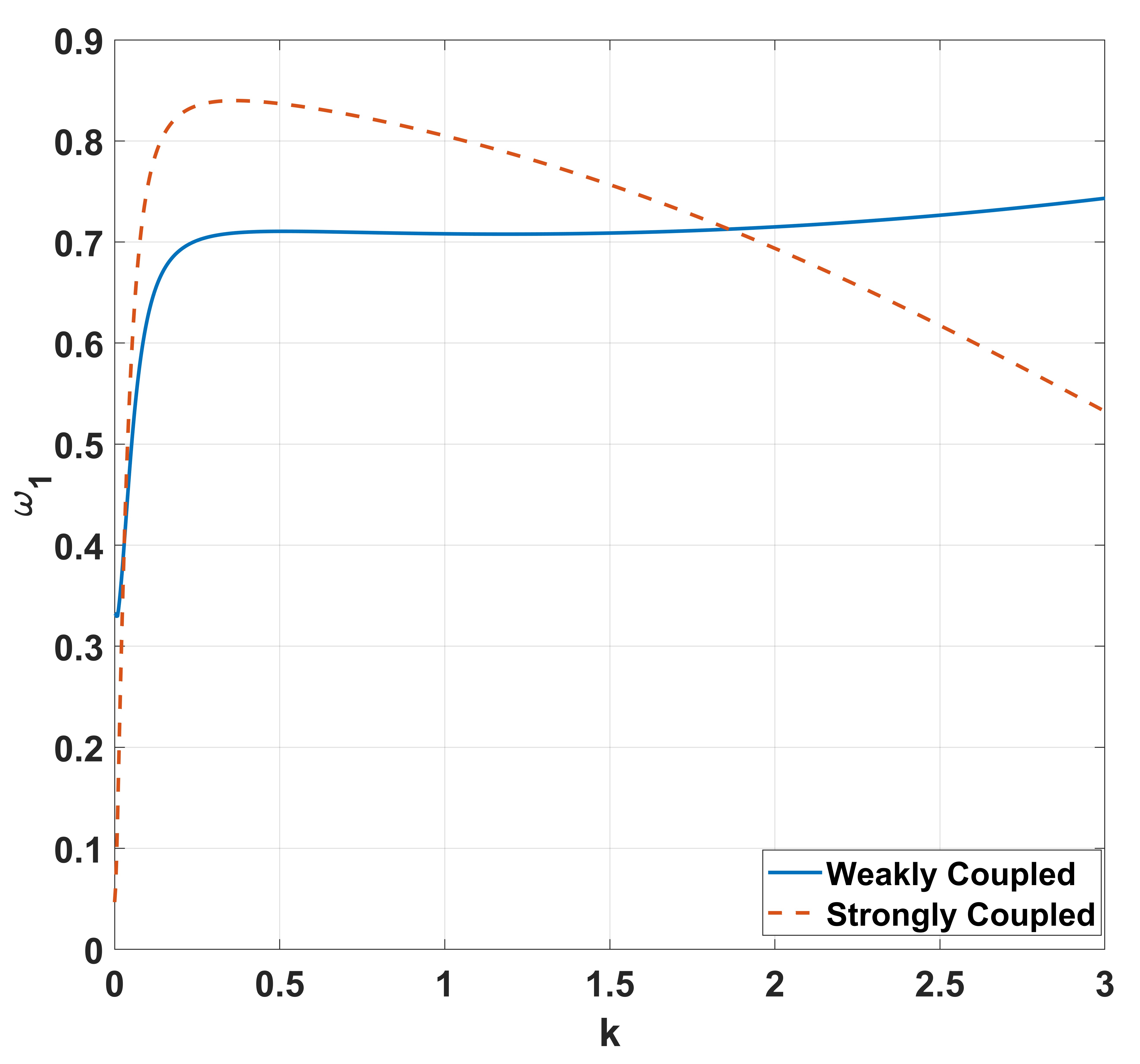}  
  \caption{}
  \label{dispersionrelationoffirstkindforweaklyandstronglycoupledplasma230523}
\end{subfigure}
\begin{subfigure}{.5\textwidth}
  \centering
  \includegraphics[width=.8\linewidth]{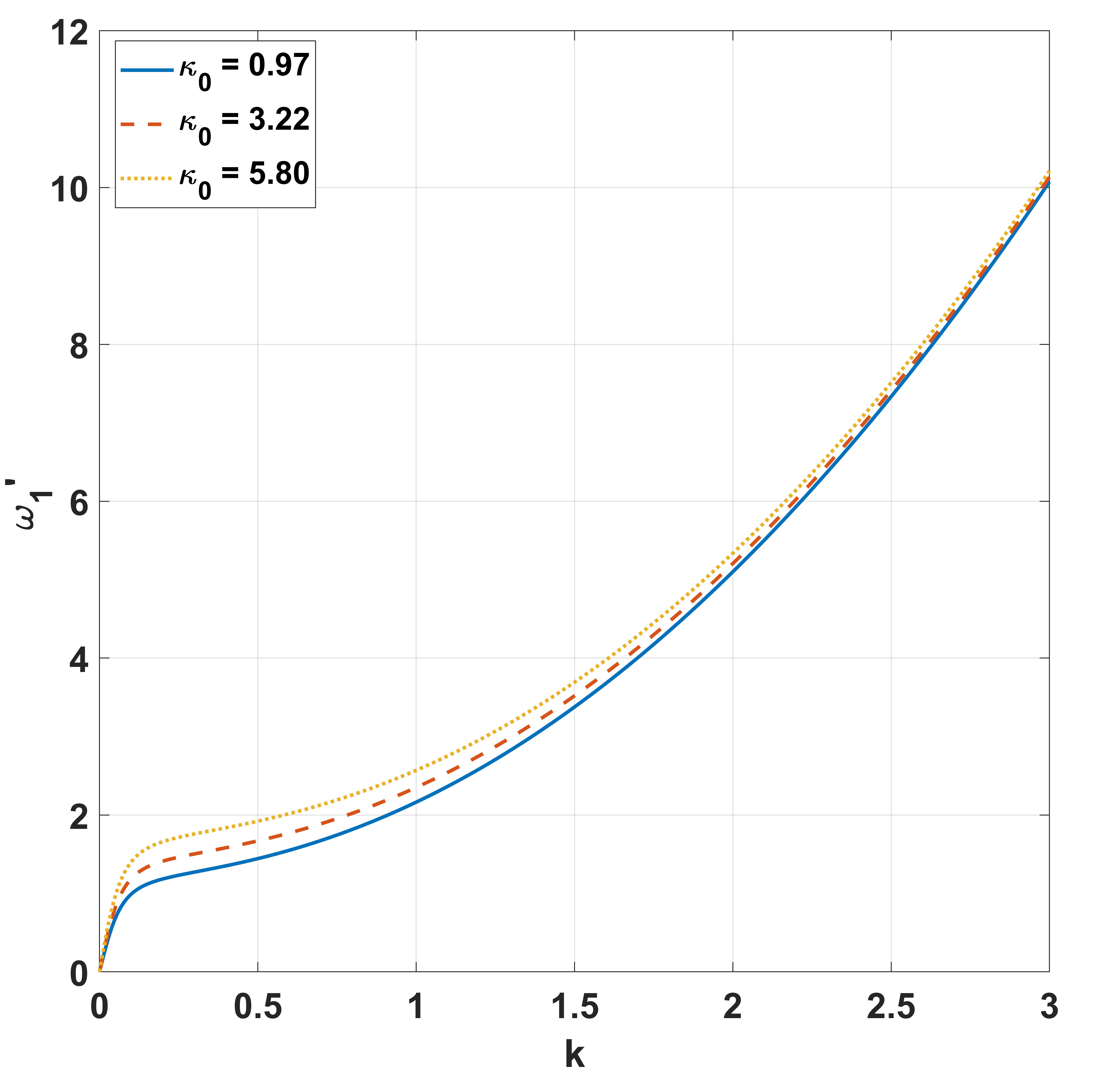}  
  \caption{}
  \label{Dispersionofsecondkindfordifferentkappa0}
\end{subfigure}
\caption{(a) Dispersion relation of the first kind for weak and strong coupling (b) Dispersion relation for second for different lattice parameter $\kappa_0$ at $T_{*0}$ }
\label{fig3}
\end{figure}

Here, at first, we will discuss the advantage of equation (\ref{eq2}) over equation (\ref{eq25}). The dispersion relation arises with the equation (\ref{eq2}) is (\ref{eq10}) and the dispersion relation arises with the help of the equation (\ref{eq25}) is equation (\ref{eq29}). Clearly, the equation (\ref{eq10}) has three roots which are shown in equation (\ref{eq13}) among them two roots $\omega_1$ and $\omega_2$ are the stable roots. Figure (\ref{dispersionrelationoffirstkindforroots230523}) represents these two stables roots for strong coupling region. From these two roots, $\omega_1$ can be stated as the ``slow" mode whereas $\omega_2$ can be described as the ``fast" mode of the dust acoustic waves. Now. if we give attention to figures (\ref{Dispersionofsecondkindwithandwithoutcoupling}),(\ref{Dispersionofsecondkindfordifferenteta0}),(\ref{Dispersionofsecondkindfordifferentuu0}), and (\ref{Dispersionofsecondkindfordifferentkappa0}) we can see that there is only one mode is there. Because between the two roots of the equation \ref{eq29} only one root ${\omega_1}'$ is the stable one. 
\par
Now, if we compare the three figures of (\ref{fig1}), where figure (\ref{dispersionrelationoffirstkindforroots230523}) arises from equation (\ref{eq13}) and the rest of the two figures (\ref{Dispersionofsecondkindwithandwithoutcoupling}) and (\ref{Dispersionofsecondkindfordifferenteta0}) emerge from equation \ref{eq29}, we can see that the slow mode of figure (\ref{dispersionrelationoffirstkindforroots230523}) has the exact same nature as that of the rest two figures. From this, we can clearly say that the stable root of equation (\ref{eq29}) represents only the ``slow" mode. Again only one curve, i.e., ``without coupling" curve of the figure (\ref{Dispersionofsecondkindwithandwithoutcoupling}) did not have any negative slope but all the other curves of figure \ref{fig1} including ``with coupling" curve of the figure (\ref{Dispersionofsecondkindwithandwithoutcoupling}) have some region with negative slope. In the figure (\ref{Dispersionofsecondkindfordifferenteta0}) the dispersion relation has been shown for different values of the dust longitudinal viscosity and the dispersion curves become steeper with the increase of the value of viscosity. So the probability of viscous damping becomes greater than the Landau-Vlasov damping \cite{Goswami2019}. On the other hand, the effect of the viscous coefficients on the dispersion relation (\ref{eq10}) is negligible as the value of $\tau_0$ mainly depends on the dust temperature $T_d$. The dust temperatures have a very low value in comparison to that of the other species and the value of $\tau_0$ is inversely proportional to the dust temperature.

\par
In figure \ref{fig2} we have shown the dispersion relation of two stable modes of (\ref{eq10}) and one stable mode of (\ref{eq29}) due to different streaming velocities of dust. The slow mode of equation (\ref{eq10}) (figure \ref{dispersionrelationoffirstkindfordifferentu02230523}) and only stable mode of equation (\ref{eq29}) (figure \ref{Dispersionofsecondkindfordifferentuu0}) have the same nature whereas the fast mode of equation (\ref{eq10}) has the opposite nature (figure \ref{dispersionrelationoffirstkindfordifferentu0230523}). By a careful look at the figures (\ref{dispersionrelationoffirstkindfordifferentu0230523}) and (\ref{dispersionrelationoffirstkindfordifferentu02230523}) we can see that the fast and slow modes intersect each other for the streaming velocity $u_0 = 0.09$.
\par
In figure (\ref{dispersionrelationoffirstkindforweaklyandstronglycoupledplasma230523}) we have shown the change in the dispersion relations due to coupling in the first case. In the strongly coupled region, the fast mode of the equation (\ref{eq10}) has the negative slope $(\frac{d\omega(k)}{dk}<0)$. In figure (\ref{Dispersionofsecondkindfordifferentkappa0}) we can see the slopes of the curves positive for most of the time and the amount of slope is increasing with the increase of the lattice parameter $(\kappa_0)$.

\subsection{Discussions on Non-linear waves:}\label{sec5.2}
\begin{figure}
\begin{subfigure}{.5\textwidth}
  \centering
  \includegraphics[width=.85\linewidth]{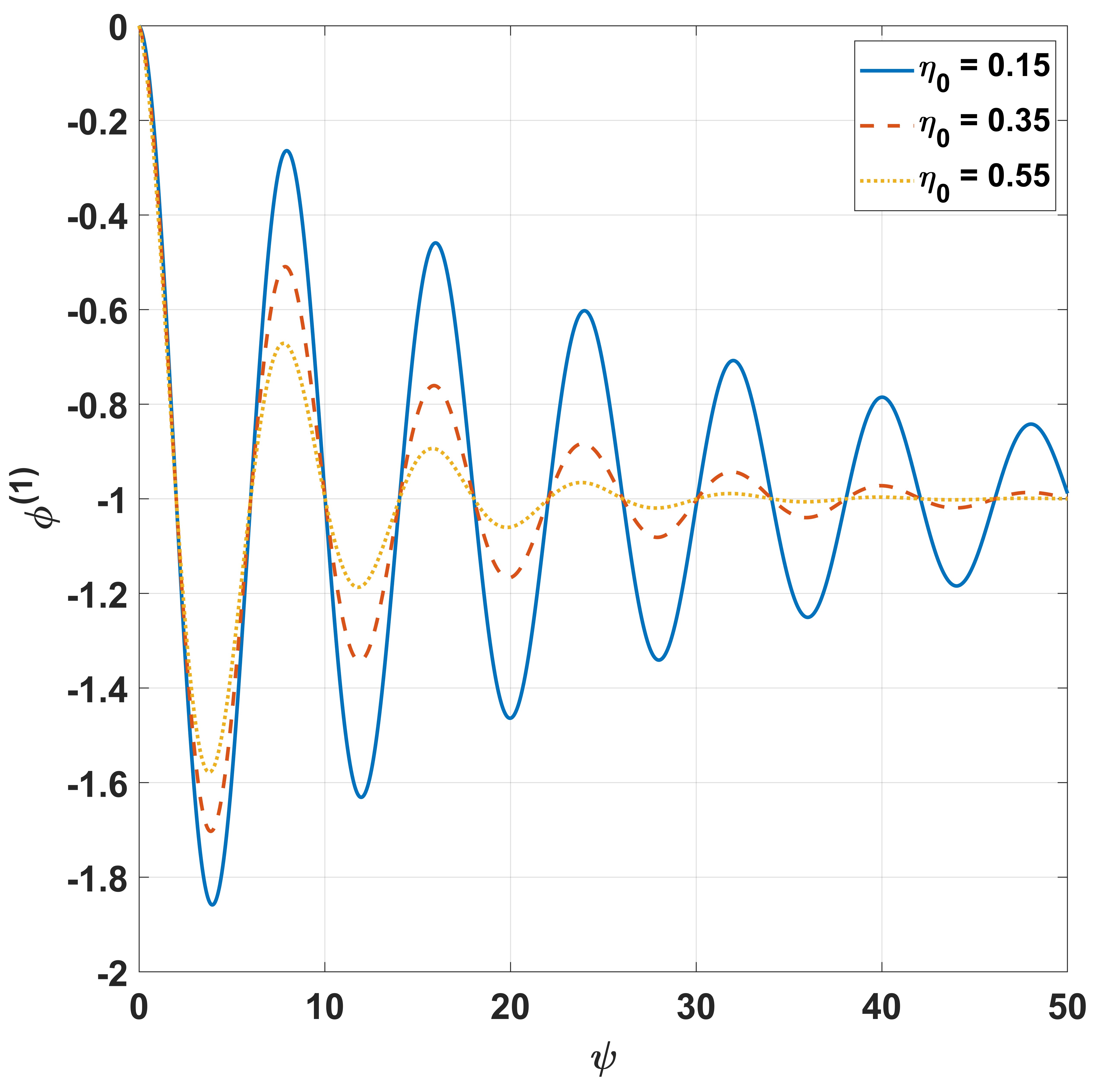}  
  \caption{}
  \label{OSWoffirstkindfordifferenteta}
\end{subfigure}
\begin{subfigure}{.5\textwidth}
  \centering
  \includegraphics[width=.85\linewidth]{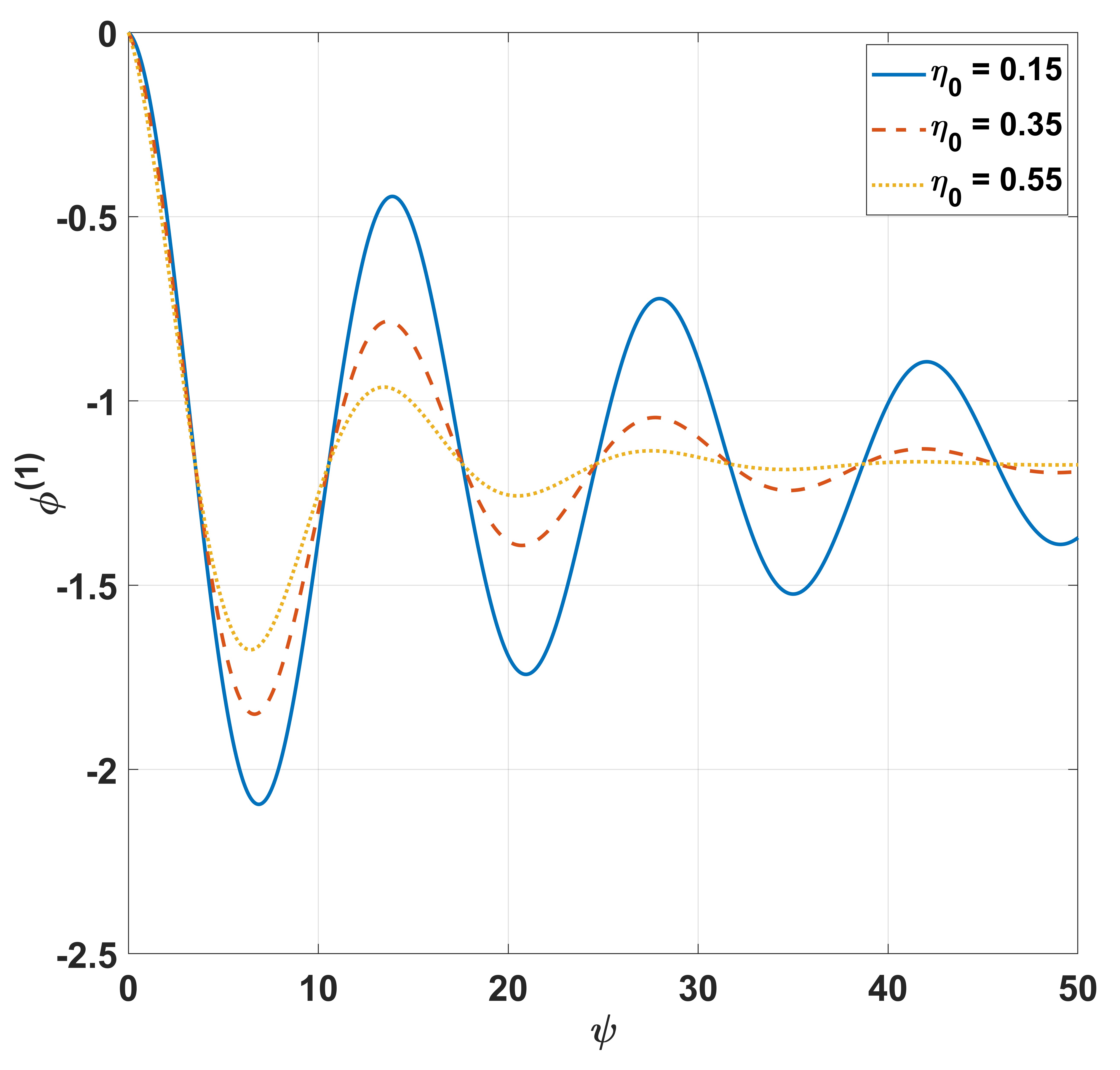}  
  \caption{}
  \label{OSWofsecondkindfordifferenteta}
\end{subfigure}
\begin{subfigure}{.5\textwidth}
  \centering
  \includegraphics[width=.85\linewidth]{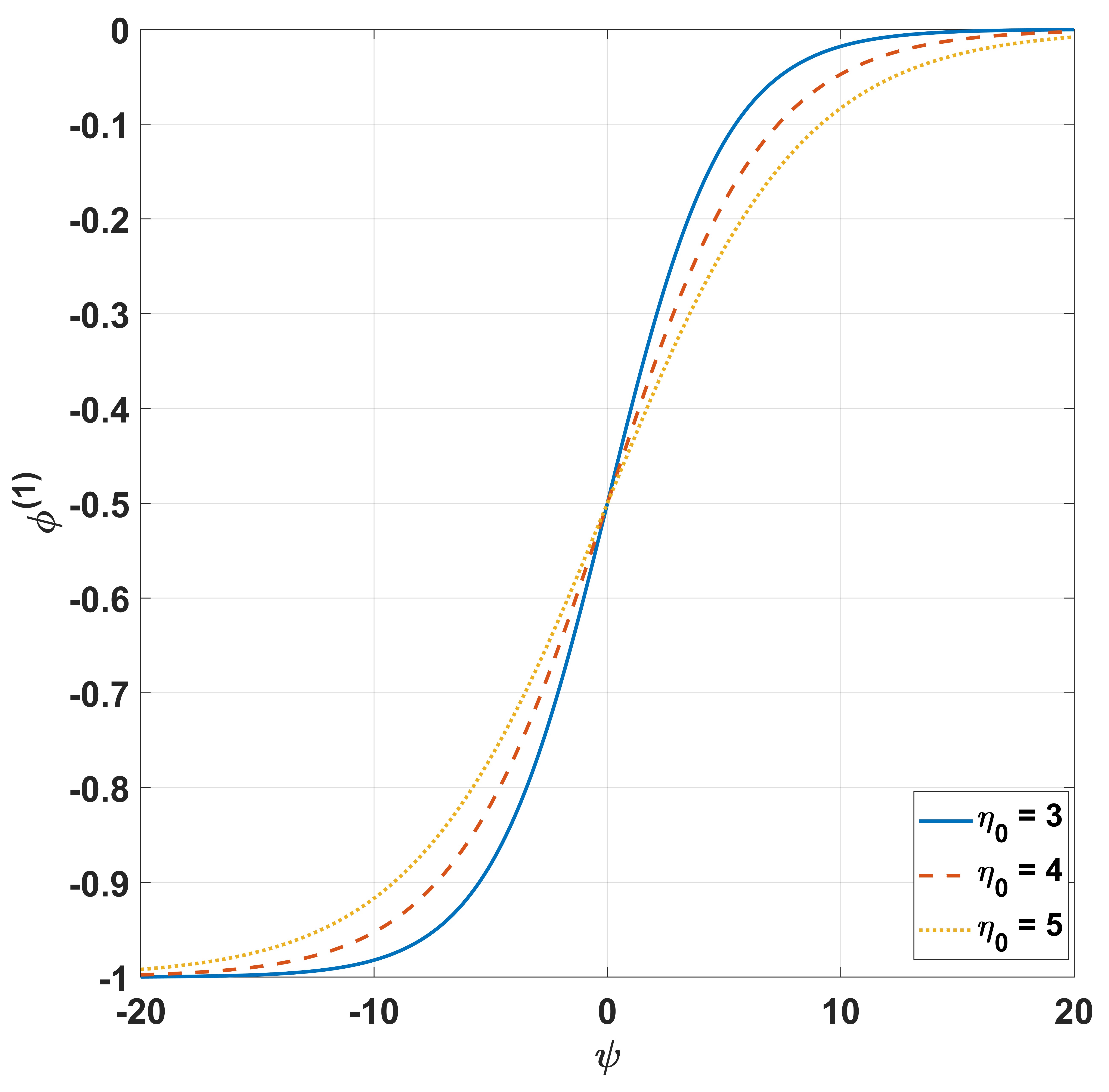}  
  \caption{}
  \label{MSWoffirstkindfordifferenteta}
\end{subfigure}
\begin{subfigure}{.5\textwidth}
  \centering
  \includegraphics[width=.85\linewidth]{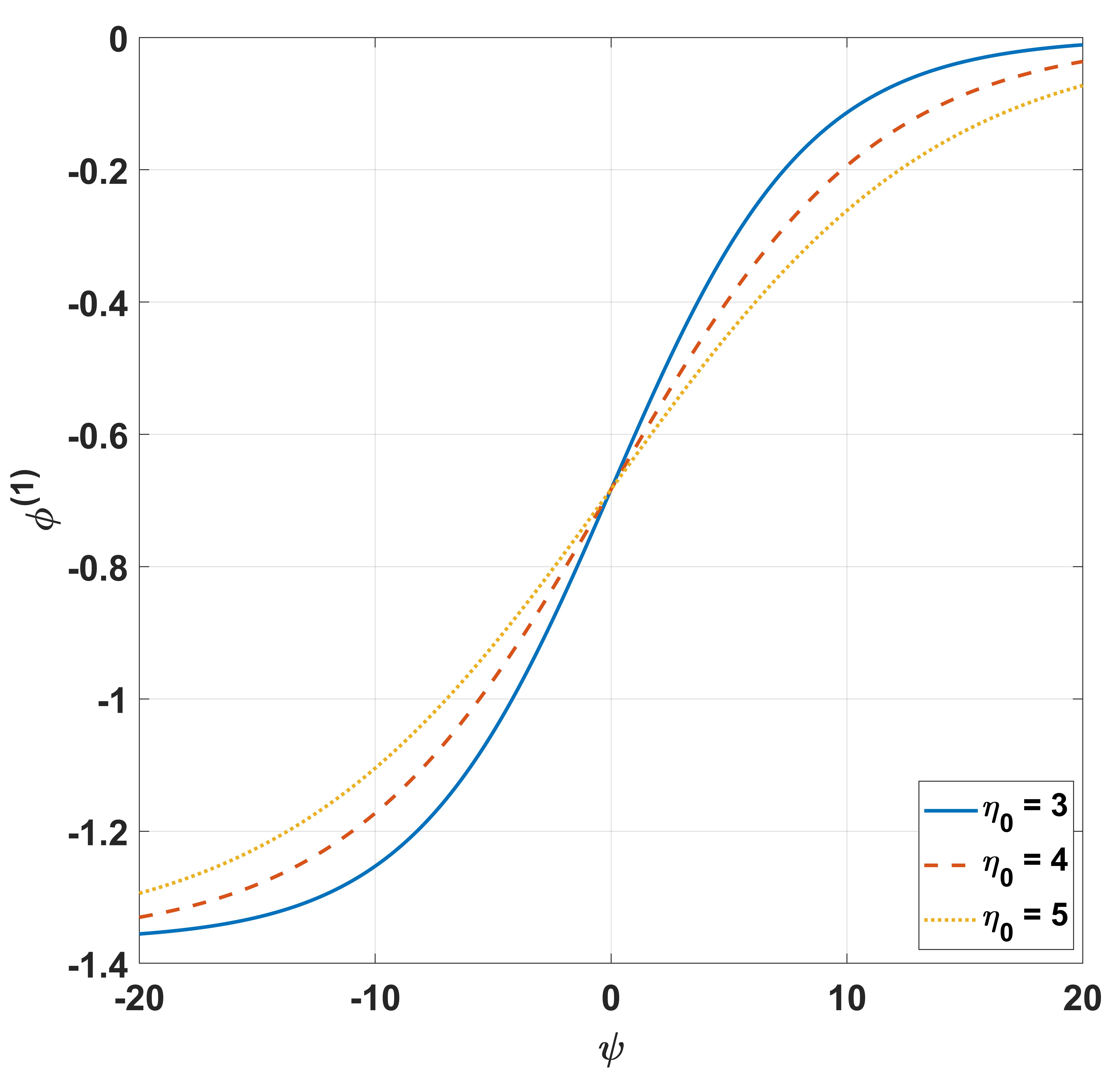}  
  \caption{}
  \label{MSWofsecondkindfordifferenteta}
\end{subfigure}
\caption{Oscillatory Shock Waves for (a) first kind (b) second kind and Monotonic Shock Waves for (c) first kind (d) second kind with different viscous coefficients}
\label{fig4}
\end{figure}

Now, we will discuss the nonlinear waves for both the treatments shown in sections  \ref{sec3.2} and \ref{sec4}. From equation (\ref{eq14}) we can see that viscoelastic relaxation time $(\tau_m)$ is used as a stretched quantity whereas from equation (\ref{eq26}) we can see that the normalized electrostatic temperature $(d)$ is treated as perturbative quantity. Both of these two considerations lead to the KdVB equation but with a very important difference. That important difference is one of the main focus of this work. If we look at the equation (\ref{eq16}) which represents the coefficients of the KdVB equation for the first kind of treatment the effect of coupling is absent as coefficients are independent of $\tau_0$. But if we look at the equation (\ref{eq34}) we can see that the coefficients have a direct dependence on coupling in terms of electrostatic temperature. The meaning of direct dependence is that in the first kind of treatment, though there is no coefficient containing viscoelastic relaxation time, there must be a dependency on coupling as there is a term of viscosity in the KdVB equation. Now we will discuss the variation of the shock waves with very important plasma parameters in this context. In figure (\ref{fig4}) we have described the effect of the longitudinal viscous coefficient on two types of shock waves. When the viscous coefficient has a small value we get the oscillatory shock waves (OSW) for both the treatments. It is quite obvious that the increase of the viscous coefficient is increasing the effect of the shock for OSW and it is shown for both the figures (\ref{OSWoffirstkindfordifferenteta}) and (\ref{OSWofsecondkindfordifferenteta}). Another important factor we want to describe here for the same two figures is (\ref{OSWoffirstkindfordifferenteta}) and (\ref{OSWofsecondkindfordifferenteta}). The wavelength of the figure (\ref{OSWoffirstkindfordifferenteta}) is greater than the wavelength of the figure (\ref{OSWofsecondkindfordifferenteta}). Also, the amplitude of the first one is a little bit less than the second one.  These two effects have an explanation that the effect of coupling for the first kind is absorbed by the rest of the parameters but the same effect arises in the second case independently. The difference due to different treatments on coupling coefficients has also been observed for the monotonic shock waves (MSWs) ( figures \ref{MSWoffirstkindfordifferenteta} and \ref{MSWofsecondkindfordifferenteta}).

\par
\begin{figure}
\begin{subfigure}{.5\textwidth}
  \centering
  \includegraphics[width=.85\linewidth]{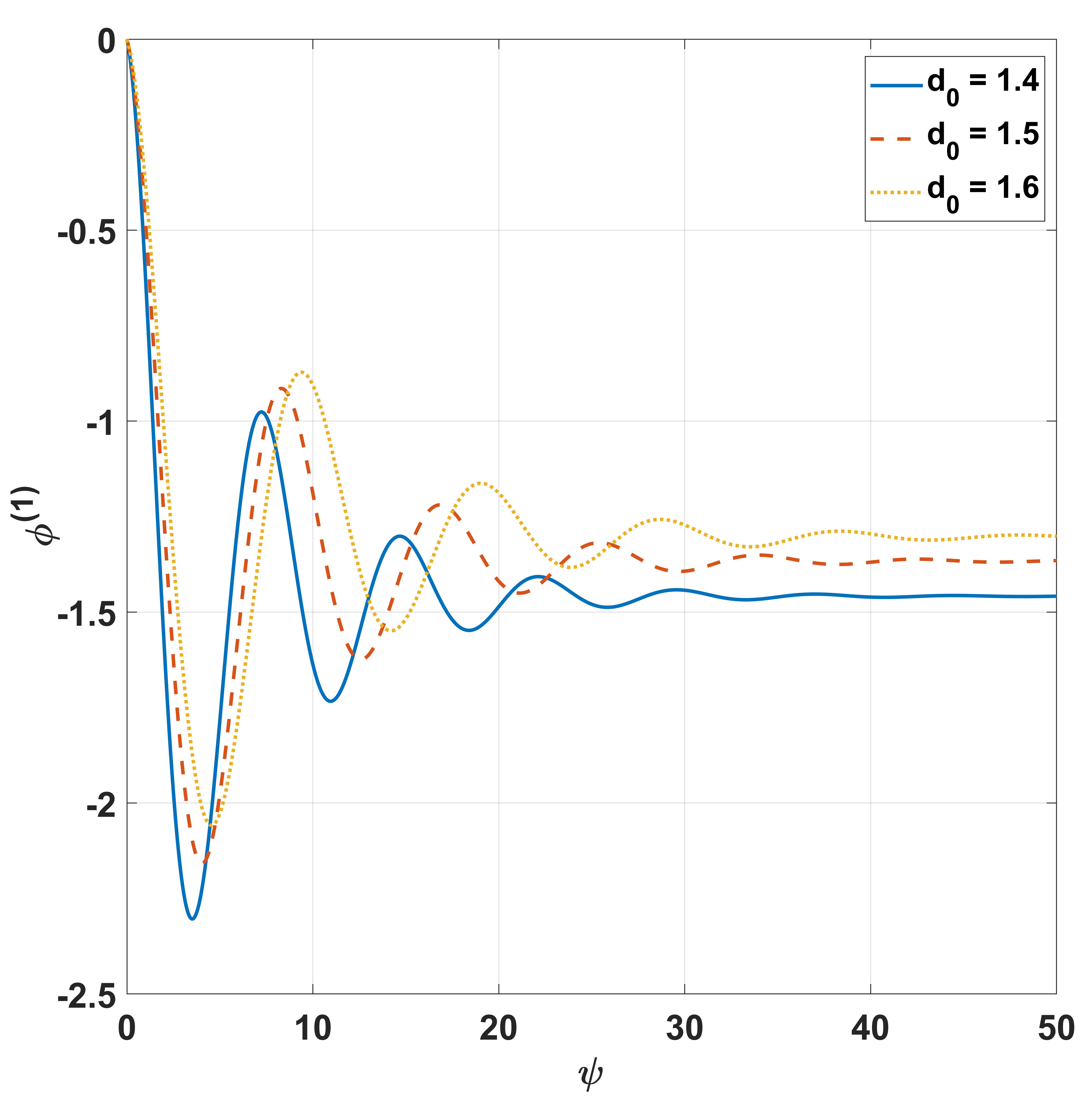}  
  \caption{}
  \label{OSWofsecondkindfordifferentd0}
\end{subfigure}
\begin{subfigure}{.5\textwidth}
  \centering
  \includegraphics[width=.85\linewidth]{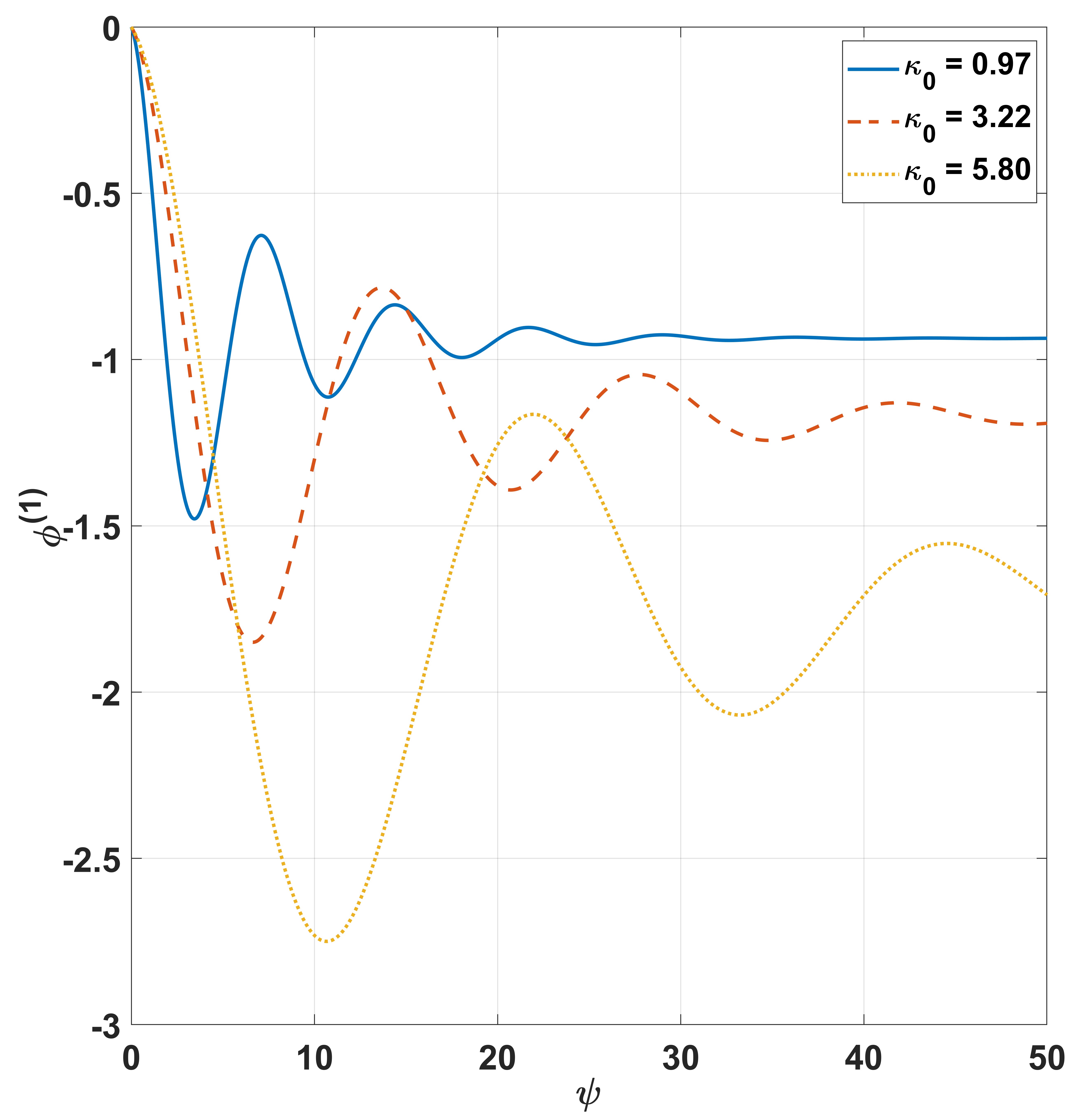}  
  \caption{}
  \label{OSWofsecondkindfordifferentkappa0}
\end{subfigure}
\begin{subfigure}{.5\textwidth}
  \centering
  \includegraphics[width=.85\linewidth]{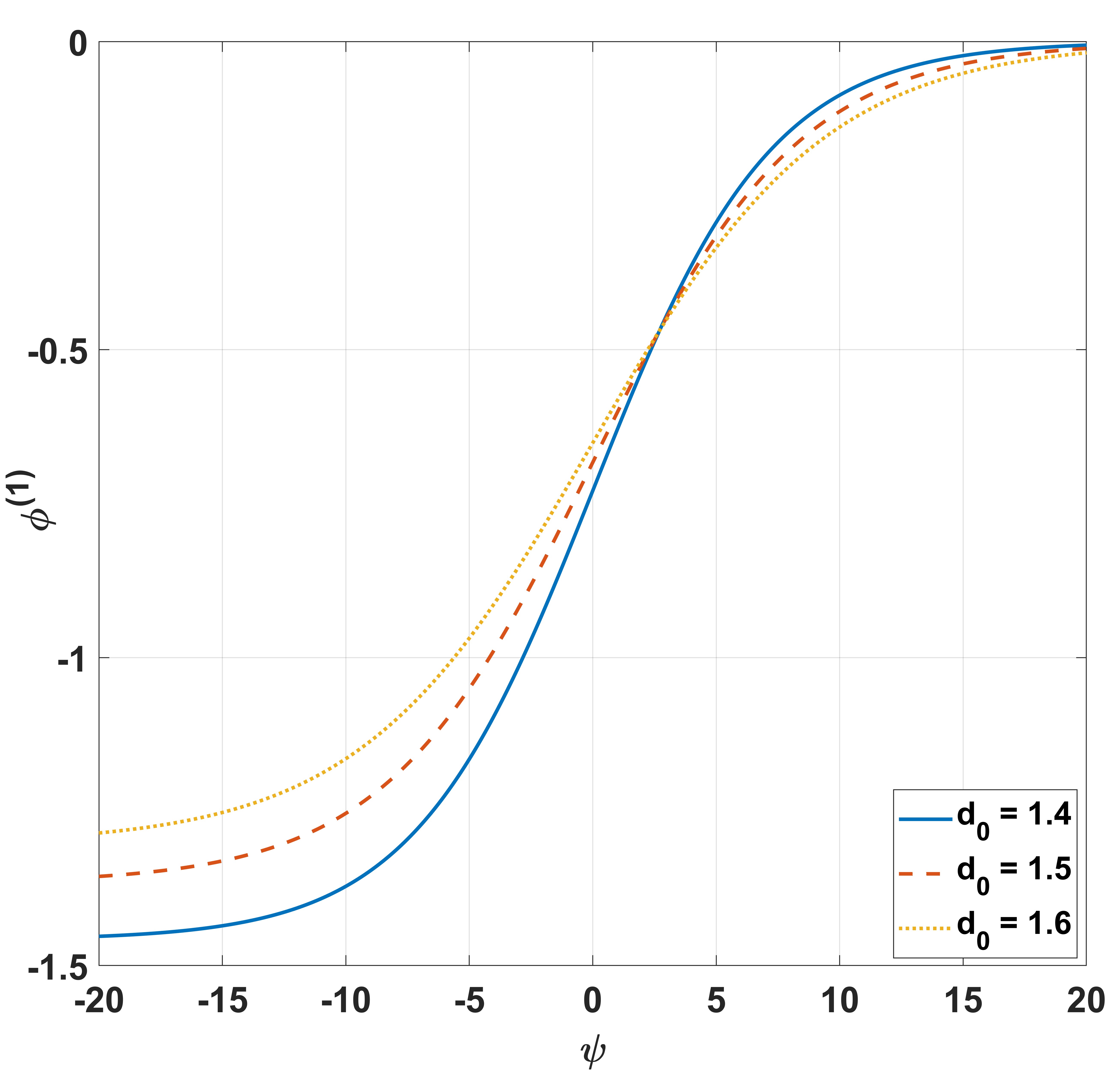}  
  \caption{}
  \label{MSWfordifferentd0secondkind}
\end{subfigure}
\begin{subfigure}{.5\textwidth}
  \centering
  \includegraphics[width=.85\linewidth]{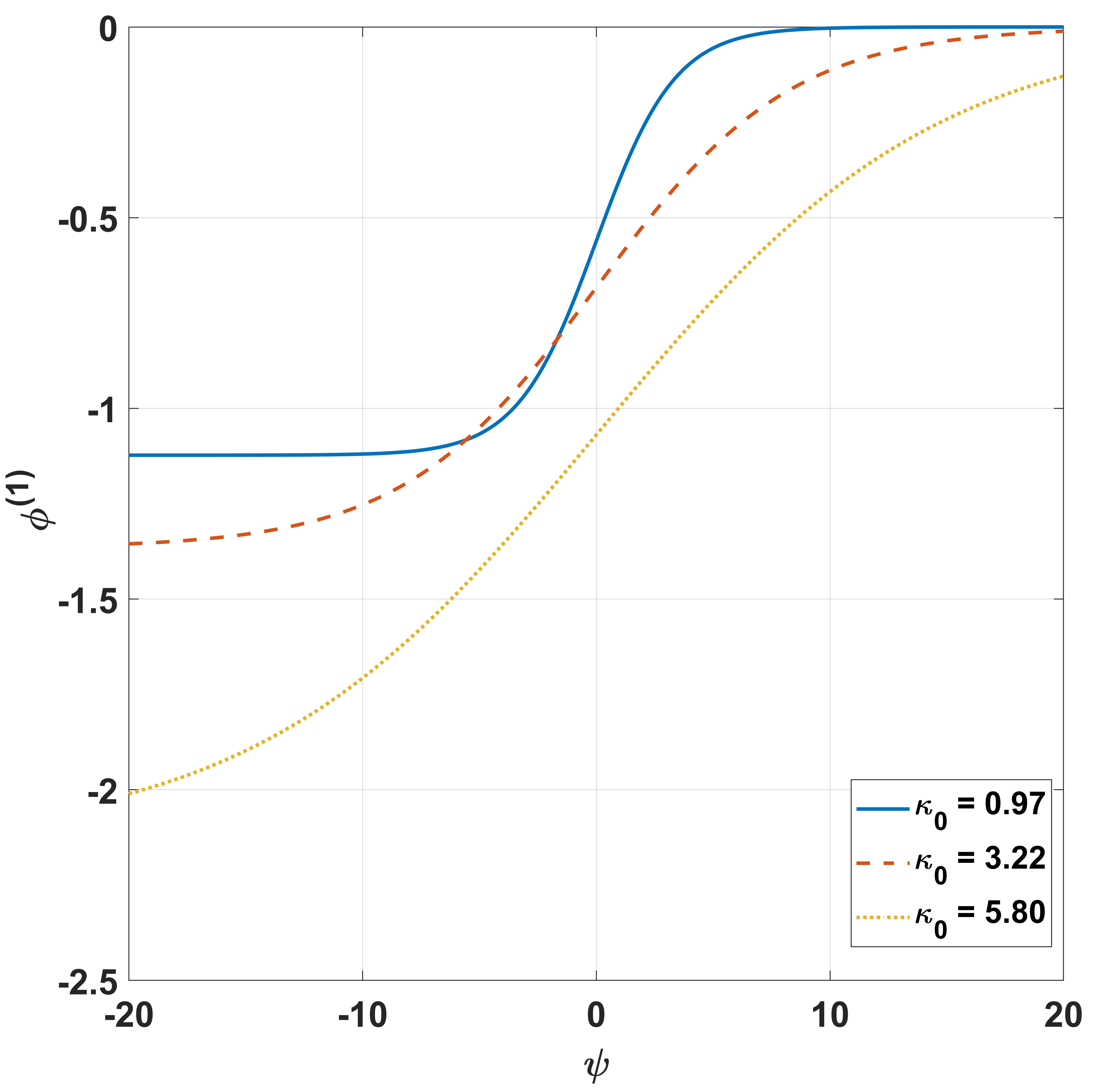}  
  \caption{}
  \label{MSWfordifferentkappa0secondkind}
\end{subfigure}
\caption{Oscillatory Shock Waves for different normalized (a) electrostatic temperature $(d_0)$ (b) lattice parameter $(\kappa_0)$ and Monotonic Shock Waves for different normalized (c) electrostatic temperature $(d_0)$ (d) lattice parameter $(\kappa_0)$}
\label{fig5}
\end{figure}

 Also, the normalized electrostatic temperature has relations with the lattice parameter $(\kappa_0)$ (Equation \ref{eq28}). In figure (\ref{fig5}) we have shown the variation of OSWs and MSWs for different $d_0$ and $\kappa_0$. With the increase of both the parameters, the amplitude and the wavelength of the OSW are increasing. The MSWs are also increasing the strength of the shock for both parameters. Physically we can say that the increase of the electrostatic temperature and lattice parameters when  all the other parameters are constant leads to a decrease in the coupling of the system and that is why the dust particles can move more freely. It is worth mentioning that the value $\kappa_0 = 5.80$ has a greater effect on both types of shocks. 

\begin{figure}
\begin{subfigure}{.5\textwidth}
  \centering
  \includegraphics[width=.85\linewidth]{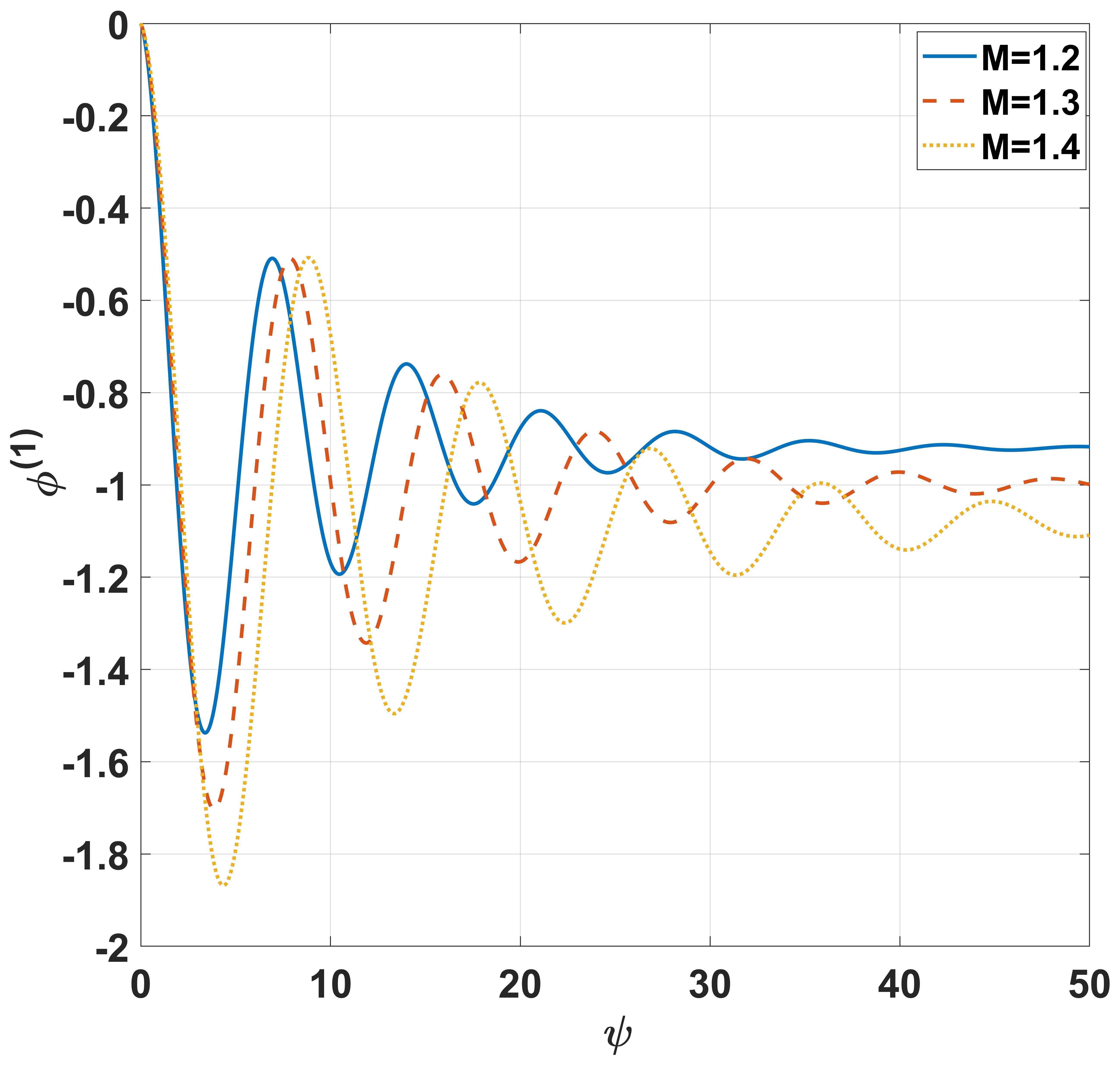}  
  \caption{}
  \label{OSWoffirstkindfordifferentM}
\end{subfigure}
\begin{subfigure}{.5\textwidth}
  \centering
  \includegraphics[width=.85\linewidth]{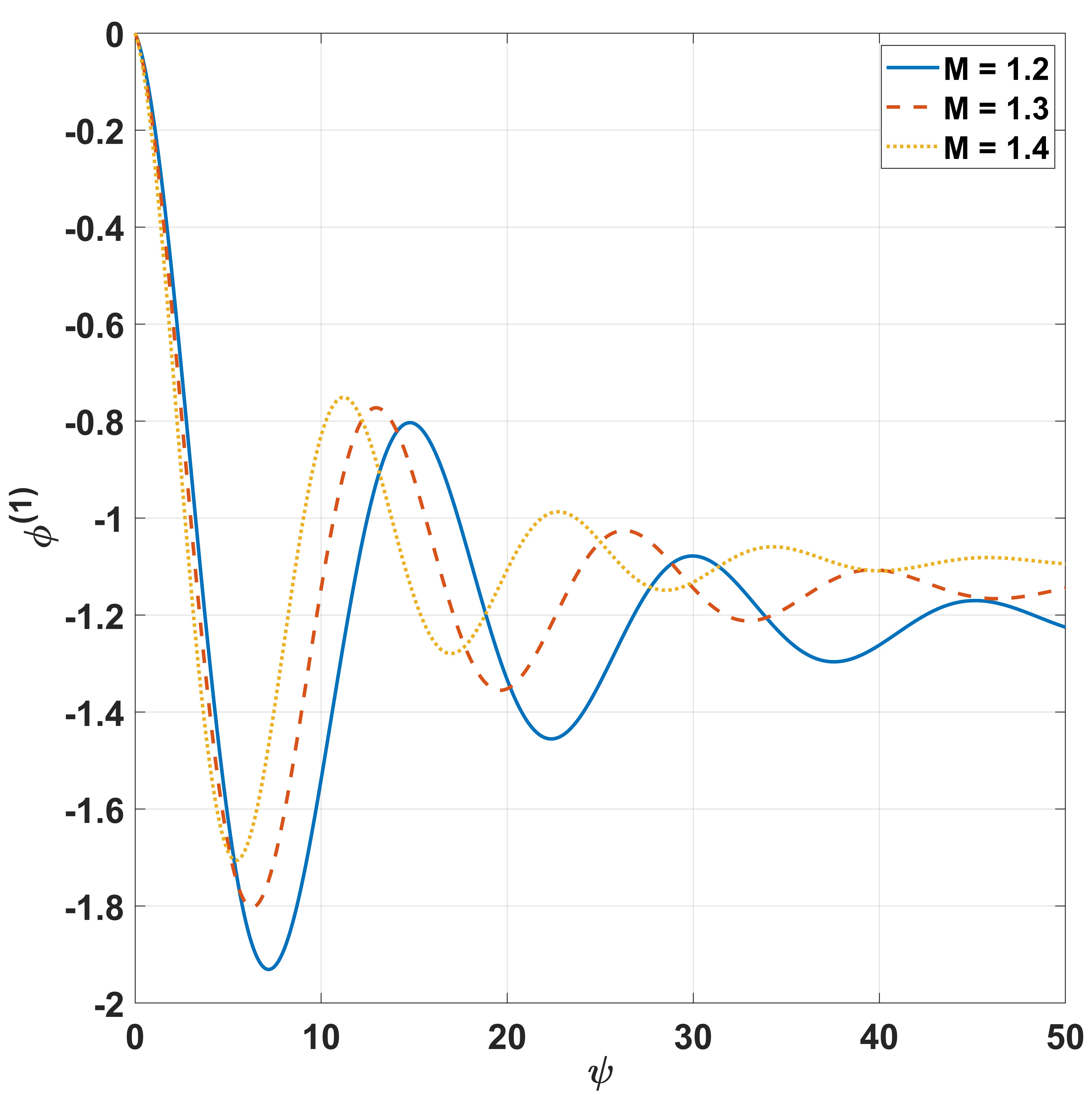}  
  \caption{}
  \label{OSWofsecondkindfordifferentM}
\end{subfigure}
\caption{Oscillatory Shock Waves for different Mach number in (a) first kind (b) second kind}
\label{fig6}
\end{figure}

\begin{figure}
\begin{subfigure}{.5\textwidth}
  \centering
  \includegraphics[width=.85\linewidth]{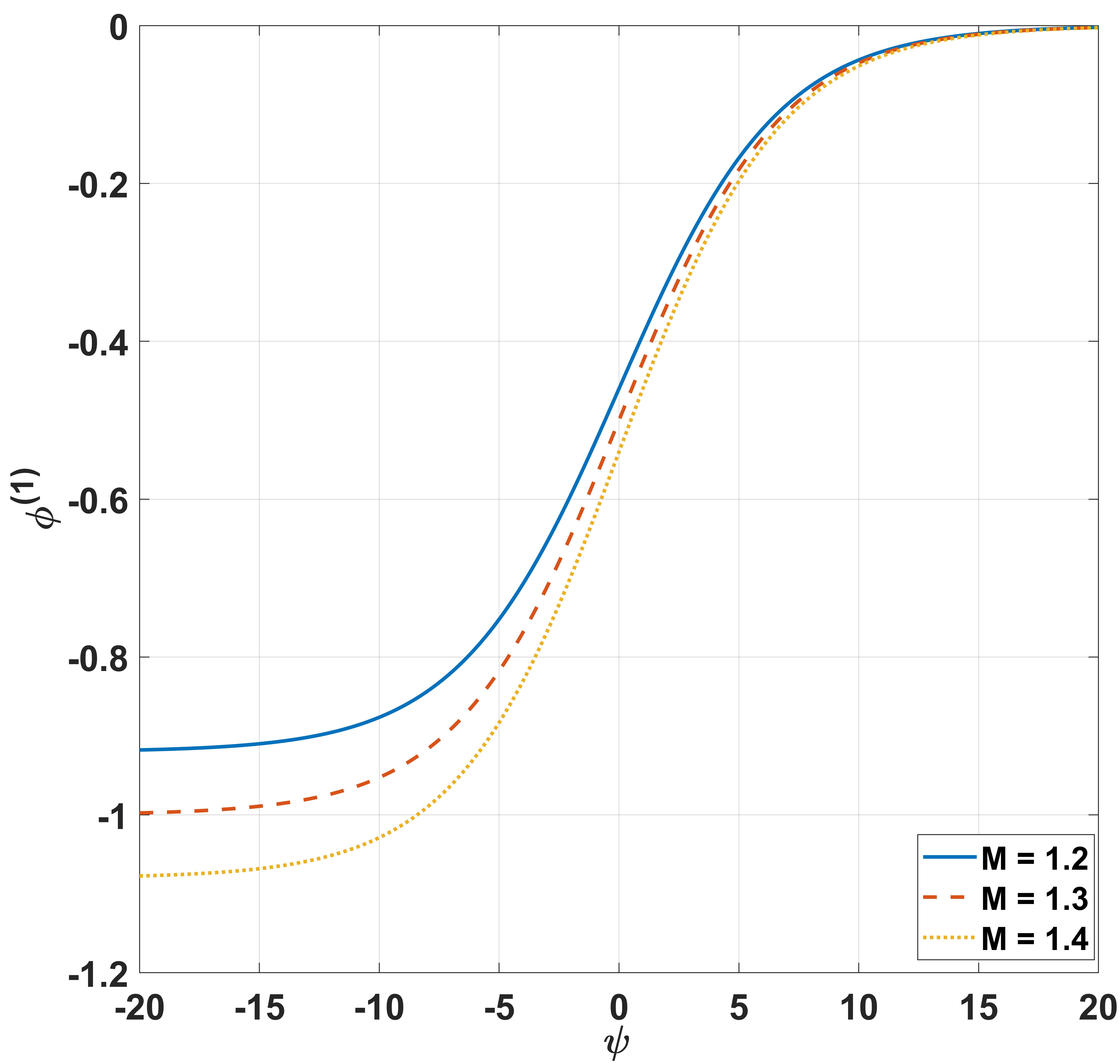}  
  \caption{}
  \label{MSWoffirstkindfordifferentM}
\end{subfigure}
\begin{subfigure}{.5\textwidth}
  \centering
  \includegraphics[width=.85\linewidth]{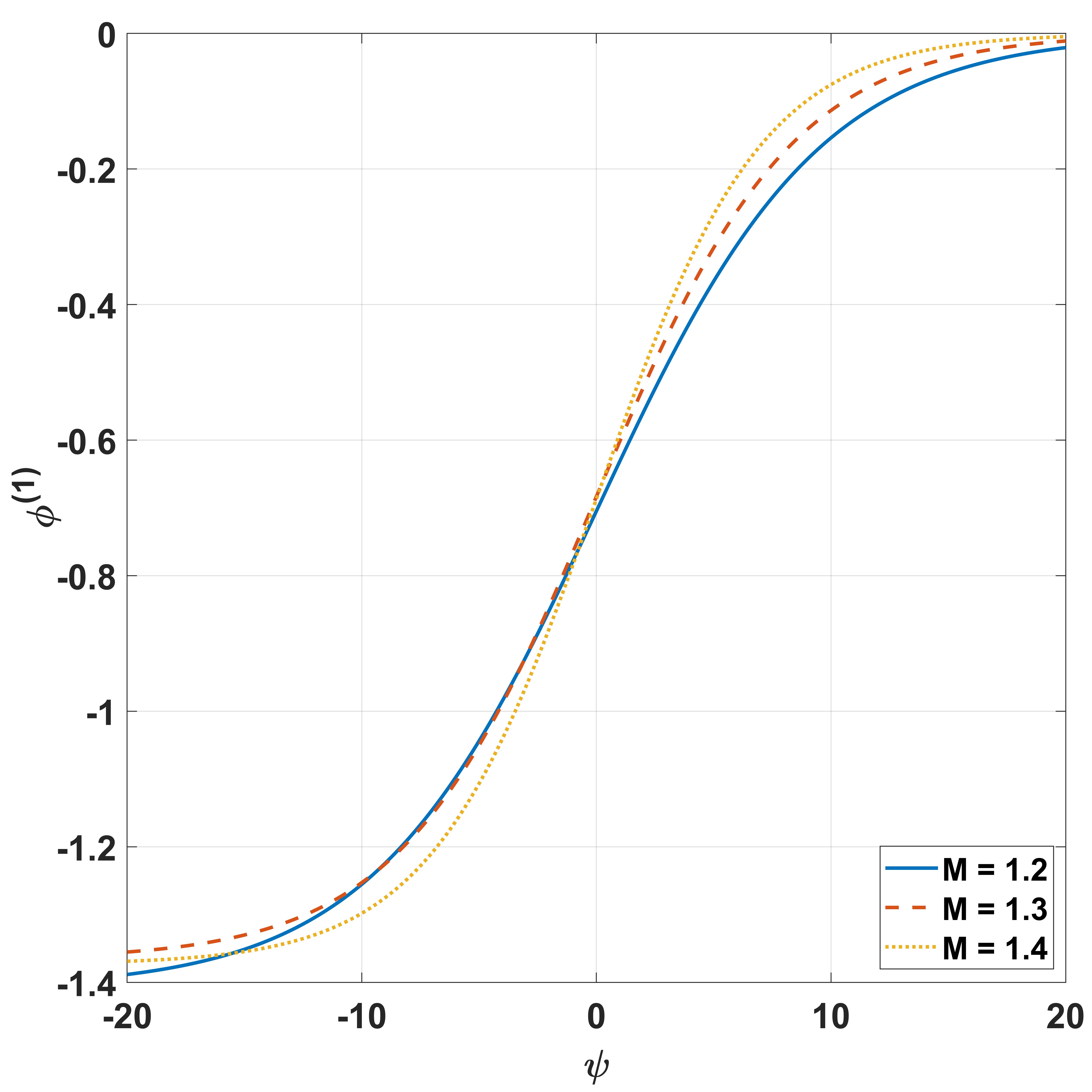}  
  \caption{}
  \label{MSWofsecondkindfordifferentM}
\end{subfigure}
\caption{Monotonic Shock Waves for different Mach number in (a) first kind (b) second kind}
\label{fig7}
\end{figure}

\begin{figure}
\begin{subfigure}{.5\textwidth}
  \centering
  \includegraphics[width=.85\linewidth]{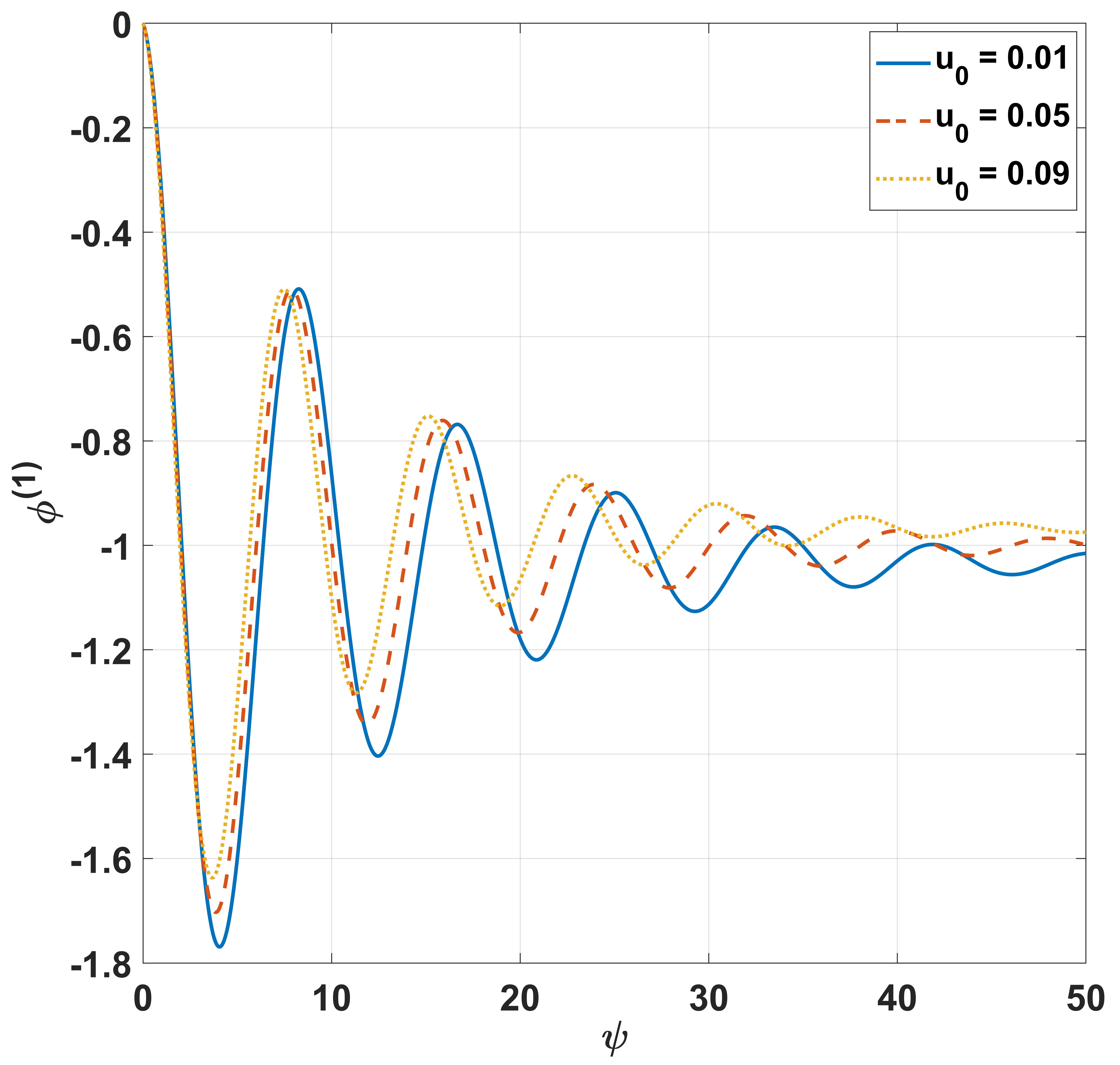}  
  \caption{}
  \label{OSWoffirstkindfordifferentstreamingvelocity}
\end{subfigure}
\begin{subfigure}{.5\textwidth}
  \centering
  \includegraphics[width=.85\linewidth]{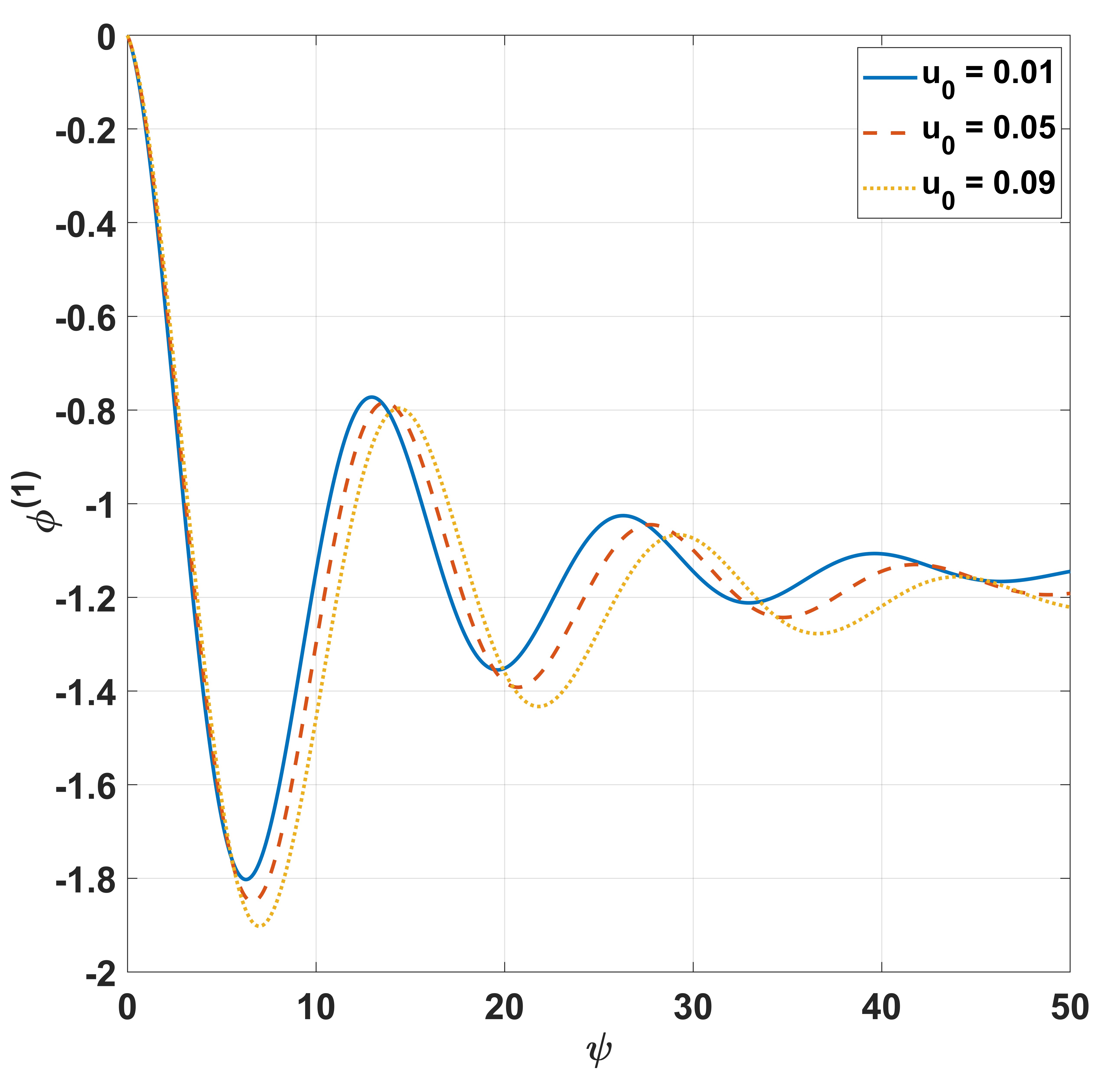}  
  \caption{}
  \label{OSWofsecondkindfordifferentstreamingvelocity}
\end{subfigure}
\caption{Oscillatory Shock Waves for different Streaming Velocity in (a) first kind (b) second kind}
\label{fig8}
\end{figure}

\begin{figure}
\begin{subfigure}{.5\textwidth}
  \centering
  \includegraphics[width=.85\linewidth]{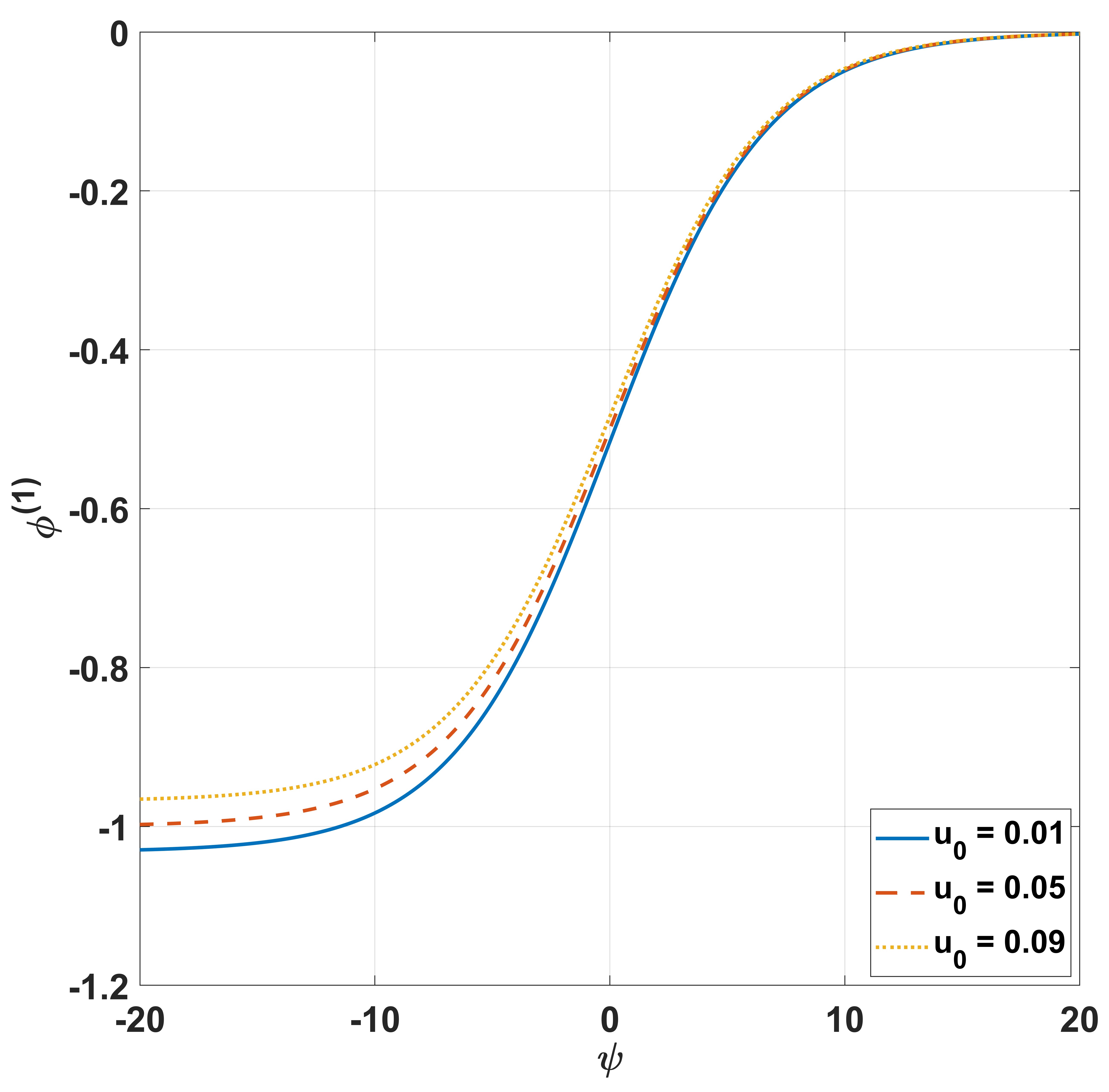}  
  \caption{}
  \label{MSWoffirstkindfordifferentstreamingvelocity}
\end{subfigure}
\begin{subfigure}{.5\textwidth}
  \centering
  \includegraphics[width=.85\linewidth]{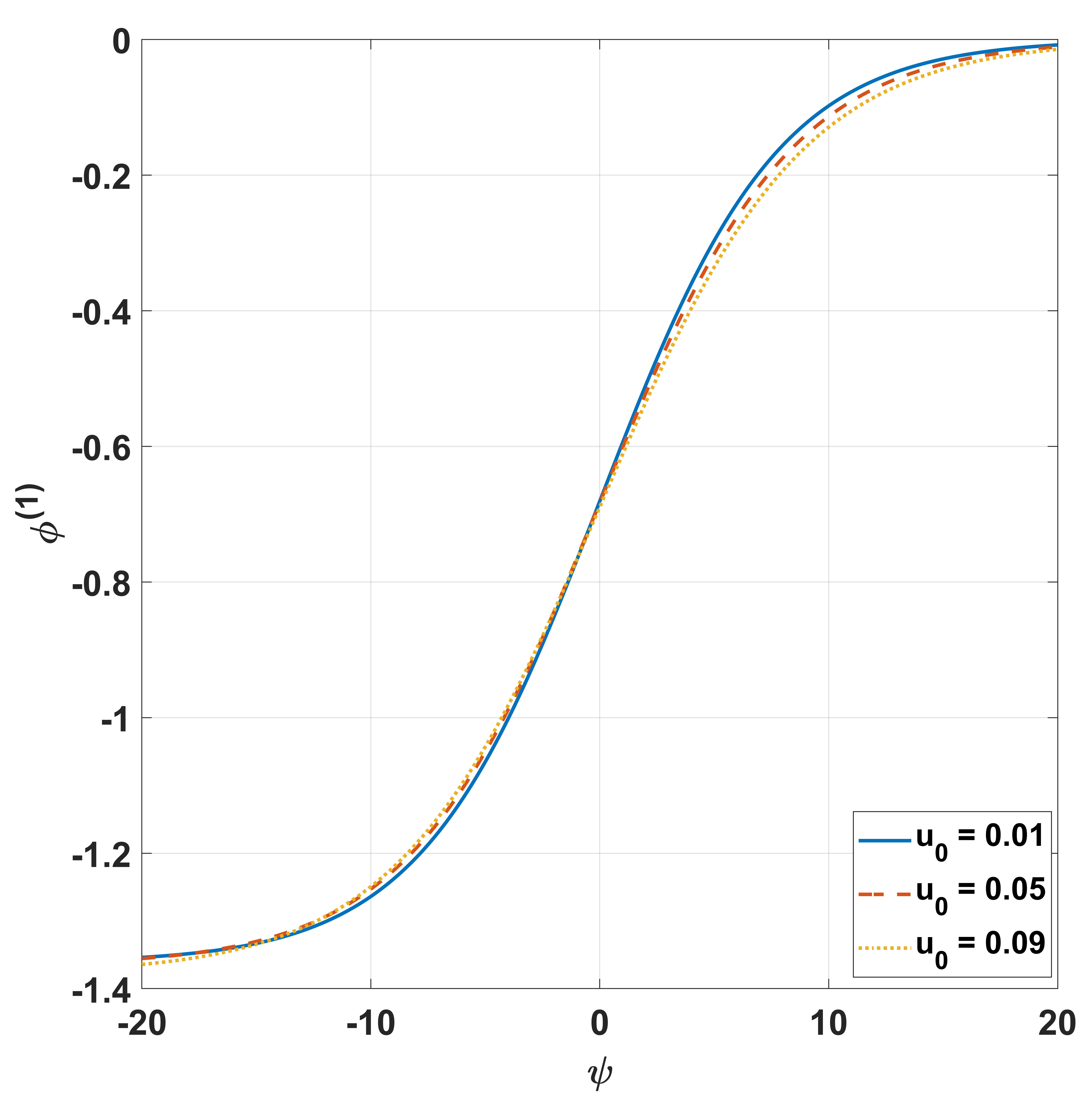}  
  \caption{}
  \label{MSWofsecondkindfordifferentstreamingvelocity}
\end{subfigure}
\caption{Monotonic Shock Waves for different Mach number in (a) first kind (b) second kind}
\label{fig9}
\end{figure}
\par
Among all the common plasma parameters of the two treatments the Mach number $(M)$ and the streaming velocity $(u_0)$ are another two very important parameters. From figure (\ref{fig6}) to figure (\ref{fig9}) we have discussed their importance. Let us first discuss the OSWs for different values of Mach numbers in both treatments. From figures (\ref{OSWoffirstkindfordifferentM}) and (\ref{OSWofsecondkindfordifferentM}) it is shown that the amplitude and wavelengths both increase for the first kind whereas the same parameters decrease for the second case. Now, if we give attention to the figures (\ref{MSWoffirstkindfordifferentM}) and (\ref{MSWofsecondkindfordifferentM}) we can observe that the amount of shock increases for the first case and decreases for the second case. Also in the figure (\ref{MSWofsecondkindfordifferentM}) there is a special phenomenon that can be seen that the curves for $M=1.3$ and $M=1.4$ cut the curve $M=1.2$ at least two times. Again figures (\ref{OSWoffirstkindfordifferentM}) and (\ref{MSWoffirstkindfordifferentM}) are analogous to the result shown in the figures (\ref{OSWoffirstkindfordifferenteta}) and (\ref{MSWofsecondkindfordifferenteta}). The same effects of the Mach number and viscous coefficient can be seen in all types of weakly coupled plasmas \cite{Goswami2019,goswami2020collision,Goswami_2023} but Mach number and the viscous coefficient have opposite responses on the shock if plasma is coupled anyway, e.g. via magnetic field \cite{Goswamiradiation}. 
\par
If we observe the OSWs (figures \ref{OSWoffirstkindfordifferentstreamingvelocity} and \ref{OSWofsecondkindfordifferentstreamingvelocity}) and MSWs (figures \ref{MSWoffirstkindfordifferentstreamingvelocity} and \ref{MSWofsecondkindfordifferentstreamingvelocity})for different streaming velocities in both the treatments we can see that they have the opposite effects than the effects due to different Mach numbers. Also, OSWs and MSWs for different dust streaming velocities of the first kind have opposite effects than the effects of viscous coefficients and this is again analogous with the effects for weakly coupled plasmas. So the most important statement we are making in this manuscript for the first time is in the form of two conditions which universally effective for all plasma systems.
\par
\subsubsection*{Condition 1:}
When we treat the strong coupling of a plasma system in the form of a viscoelastic relaxation time then all the other plasma parameters have shown their effect as if they are operating in weakly coupled plasma. Because the effect of the viscoelastic relaxation time has been absorbed by the longitudinal viscous coefficient.
\par
\subsubsection*{Condition 2:}
When we treat the strong coupling of a plasma system in the form of a normalized electrostatic temperature then all the other parameters have shown their effect as if they are operating in a strongly coupled plasma. As the coupling effect is independently present in the system.
\par
\emph{Condition 1} can only be different if any force (mainly dissipative force like `collision') is coupled with viscoelastic relaxation time itself.
\par
Now, we will discuss the effect of the rest of the plasma parameters on the OSW and MSW. As we consider the temperature of dust is $100$ and $1000$ times lesser than the temperature of ions and electrons, respectively, the effect of the different $q-$ values and densities for both treatments has a minimal effect. We have shown these effects in the supplementary file, where we can see the effect of the $q$ values of electrons have a different effect on the OSWs and MSWs than the $q$ values of positrons and ions.
\section{Conclusion:}\label{sec6}
In this paper, we have investigated the dust acoustic shock and solitary waves associated DAWs in an EPIDPM. To the best of our knowledge, no previous worker has ever tried two treatments for the coupling for the same plasma system. From the results, we can see that there are three most important plasma parameters common for both the treatments, these are Mach number $(M)$, the streaming velocity of dust $u_0$, and the longitudinal viscous coefficient $\eta_0$. Depending on the fact that the coupling term is directly associated with the longitudinal viscous coefficient or not determines the behavior of the rest of the parameters. So the first treatment in the nonlinear region is only effective when the density is very low of the plasma system and longitudinal viscous force is the only dissipative force present in the system. To encounter a high-density plasma by this treatment we have to consider another dissipative force like collision. However, the nonlinear behavior in the liquid crystal region can not be analyzed even after considering the collision or collision-like dissipative force. But from the discussion on the second treatment of the coupling, we can see the nonlinear effect in all the regions from liquid-crystal to the hydrodynamic phase of a dusty plasma system. 
\par
But there is an advantage of proceeding through the first treatment that is the linear behavior of dust acoustic waves can be better analysed by this. So the linear instabilities, e.g. two-stream instability can be understood more rigorously by the first treatment.
\par
There is one very important finding we have from the supplementary file that is given below as a note.
\par
 \emph{Note:} the nature of the $q-$ values of ions and positrons are the same and it is different from the $q-$ value of electrons.
\par
The region of our focus is the magnetosphere region of Saturn. However, we have not considered the magnetic field in the calculation, which is our future perspective. The study of different types of instabilities is another path that we can explore for strongly coupled plasma.

\section*{References:}
\bibliographystyle{unsrt}
\bibliography{thirdwork}
\appendix
\section{}\label{A}
The dispersion relation is given in equation \ref{eq10}. These we substitute,
\begin{equation}\label{A1}
    \omega=y-\frac{b}{3a}
\end{equation}
Therefore,
\begin{equation}
    a\left(y-\frac{b}{3a}\right)^3+b\left(y-\frac{b}{3a}\right)^2+c\left(y-\frac{b}{3a}\right)+d=0
\end{equation}
Rearranging this we get,
\begin{equation}
    y^3+\left(\frac{3ac-b^2}{3a^2}\right)y+\frac{2b^3-9abc+27a^2d}{27a^3}=0
\end{equation}
Putting $Q=\frac{3ac-b^2}{9a^2}$ and $R=\frac{2b^3-9abc+27a^2d}{54a^3}$ we get,
\begin{equation}\label{A4}
    y^3+3Qy-2R=0
\end{equation}
Considering the identity $\left(S+T\right)^3-3ST\left(S+T\right)-\left(S^3+T^3\right)=0$ and matching with the equation \ref{A4} we get,
\begin{equation}
    \left. \begin{array}{ll}
\displaystyle y=S+T
  \\[8pt]
  \displaystyle ST=-Q \\[8pt]
  \displaystyle S^3+T^3=2R \\[8pt]
 \end{array}\right\}
\end{equation}
Now, proceed with this, and using Vieta's formula we get the value of $S$ and $T$ as,
\begin{equation}
    \left. \begin{array}{ll}
\displaystyle S=\sqrt[3]{R+\sqrt{R^2+Q^3}}
  \\[8pt]
  \displaystyle T=\sqrt[3]{R-\sqrt{R^2+Q^3}} \\[8pt]
 \end{array}\right\}
\end{equation}
Again with some algebraic steps, we get the following solution,
\begin{equation}\label{A7}
    \left. \begin{array}{ll}
\displaystyle y_1=S+T
  \\[8pt]
  \displaystyle y_2=\frac{-\left(S+T\right)+i\sqrt{3}\left(S-T\right)}{2} \\[8pt]
  \displaystyle y_3=\frac{-\left(S+T\right)-i\sqrt{3}\left(S-T\right)}{2} \\[8pt]
 \end{array}\right\}
\end{equation}
Again, from equations \ref{A1} and \ref{A7} we got the solution given in equation \ref{eq13}.
\section{}\label{B}
The $c_2$ and $c_3$ in the equation \ref{eq28} are given below,
\begin{eqnarray*}
    c_2&= \\
    & \frac{1}{8}\Bigg[\left(1+q_e\right)\left(3-q_e\right)\alpha_e {\sigma_e}^2-\left(1+q_i\right)\left(3-q_i\right)\alpha_i {\sigma_i}^2-\left(1+q_p\right)\left(3-q_p\right)\alpha_p {\sigma_p}^2\Bigg]
\end{eqnarray*}
and
\begin{eqnarray*}
   c_3 & =  \frac{1}{48}\Bigg[\left(1+q_e\right)\left(3-q_e\right) \left(5-3q_e\right)\alpha_e {\sigma_e}^3-\left(1+q_i\right)\left(3-q_i\right)\left(5-3q_i\right)\alpha_i {\sigma_i}^3 \\
    & -\left(1+q_p\right)\left(3-q_p\right)\left(5-3q_p\right)\alpha_p {\sigma_p}^3\Bigg] 
\end{eqnarray*}

\end{document}